# Current and future roles of artificial intelligence in retinopathy of prematurity

Ali Jafarizadeh, Shadi Farabi Maleki, Parnia Pouya, Navid Sobhi, Mirsaeed Abdollahi, Siamak Pedrammehr, Chee Peng Lim, Houshyar Asadi, Roohallah Alizadehsani, Ru-San Tan, Sheikh Mohammad Shariful Islam, U. Rajendra Acharya

*Abstract*— **Retinopathy of prematurity (ROP) is a severe condition affecting premature infants, leading to abnormal retinal blood vessel growth, retinal detachment, and potential blindness. While semi-automated systems have been used in the past to diagnose ROP-related plus disease by quantifying retinal vessel features, traditional machine learning (ML) models face challenges like accuracy and overfitting. Recent advancements in deep learning (DL), especially convolutional neural networks (CNNs), have significantly improved ROP detection and classification. The i-ROP deep learning (i-ROP-DL) system also shows promise in detecting plus disease, offering reliable ROP diagnosis potential. This research comprehensively examines the contemporary progress and challenges associated with using retinal imaging and artificial intelligence (AI) to detect ROP, offering valuable insights that can guide further investigation in this domain. Based on 89 original studies in this field (out of 1487 studies that were comprehensively reviewed), we concluded that traditional methods for ROP diagnosis suffer from subjectivity and manual analysis, leading to inconsistent clinical decisions. AI holds great promise for improving ROP management. This review explores AI's potential in ROP detection, classification, diagnosis, and prognosis.**

*Index Terms*— **Artificial intelligence, Convolutional neural networks, Deep Learning, Machine Learning, Ophthalmology, Retinopathy, Retinopathy of prematurity, ROPtool**

## I. INTRODUCTION

RETINOPATHY OF PREMATURITY (ROP), a proliferative vascular disorder of the retina afflicting premature and low-birth-weight infants that is linked to oxygen therapy-induced injury, can lead to retinal detachment and permanent blindness in the absence of timely diagnosis and treatment [1]. Its incidence increases with the extent of immaturity and intrauterine growth retardation [2]. Secular improvements in survival rates of preterm newborns in neonatal intensive care units combined with insufficient resources to monitor oxygen therapy in resource-limited countries have led to a rise in the number of preterm infants suffering from ROP, especially in developing countries [3]. ROP is one of the most prevalent preventable causes of blindness [4]. Every year, approximately 50,000 infants suffer permanent blindness due to ROP, with the highest disease burden seen in countries in Latin America, Southeast Asia, and Eastern Europe [5]

### A. Pathogenesis of ROP

ROP is characterized by abnormal blood vessel development in the retina with excessive vascularization [6] that is strongly linked to intensive oxygen therapy. Other contributing factors have also been proposed as pathogenetic mechanisms [7]—e.g., reduced ghrelin, oxidative stress, and erythropoietin dysregulation—the roles remain fully clarified [8]. ROP is a biphasic disease: in Phase I, administration of oxygen therapy induces cellular level hypoxia and delayed vascular growth in the retina; and in Phase II, cessation of oxygen therapy, neovascularization of retinal vessels into the vitreous body [9]. In the retina, initial iatrogenic hyperoxia suppresses the expression of vascular endothelial growth factor (VEGF), which subsequently rises to supernormal levels once oxygen therapy is withdrawn [10]. Non-oxygen-dependent factors also modulate the development of ROP. For instance, lower serum insulin-like growth factor I (IGF-I) levels were found in infants who develop ROP than those from matched controls [4].

### B. ROP in the technology age

Timely diagnosis and treatment of ROP are imperative [11]. ROP is traditionally diagnosed by ophthalmoscopic examination, which is challenging to perform in infants and requires expertise. Digital retinal images captured by RetCam,

Ali Jafarizadeh, Shadi Farabi Maleki, Navid Sobhi and Mirsaeed Abdollahi are with the Nikookari Eye Center, Tabriz University of Medical Sciences, Tabriz, Iran.

Parnia Pouya is with the Student Research Committee, Tabriz University of Medical Sciences, Tabriz, Iran.

Siamak Pedrammehr is with the Faculty of Design, Tabriz Islamic Art University, Tabriz, Iran.

Chee Peng Lim, Houshyar Asadi, and Roohallah Alizadehsani are with the Institute for Intelligent Systems Research and Innovation, Deakin University, VIC 3216, Australia.

Ru-San Tan is with the National Heart Centre Singapore, Singapore and also with the Duke-NUS Medical School, Singapore.

Sheikh Mohammed Shariful Islam is with the Institute for Physical Activity and Nutrition, School of Exercise and Nutrition Sciences, Deakin University, Geelong, VIC, Australia and also with the Cardiovascular Division, The George Institute for Global Health, Newtown, Australia and also with the Sydney Medical School, University of Sydney, Camperdown, Australia.

U. Rajendra Acharya is with the School of Mathematics, Physics, and Computing, University of Southern Queensland, Springfield, QLD 4300, Australia and also with the Centre for Health Research, University of Southern Queensland, Australia.

Ali Jafarizadeh, Shadi Farabi Maleki, and Parnia Pouya are contributed equally to this work

Corresponding authors: Ali Jafarizadeh (ali.jafarizadeh.md@gmail.com) and Siamak Pedrammehr (S.pedrammehr@tabriziau.ac.ir)

NM-200D, and smartphone-based fundus cameras provide reproducible inputs for offline analysis by experts, which may improve diagnostic performance [12-14]. Artificial intelligence (AI)-enabled analysis of retinal images is widely applied in ophthalmology for diagnosis, health monitoring, drug development, and management of diverse conditions, including diabetic retinopathy, glaucoma, age-related macular degeneration, and ROP [15-17]. In this study, we will review the state-of-the-art AI applications in ROP—including computer-based systems like ROPTool, Retinal Image Multiscale Analysis (RISA), VesselMap, Computer-Aided Image Analysis of the Retina (CAIAR), as well as newer deep learning (DL) models like iROP-DL and DeepROP—that have transformed the screening and early diagnosis of ROP. This study complements and is distinguished from prior reviews by the comprehensiveness of its examination of machine learning (ML) and DL in computer-based assisted image analysis (CBIA) of retinal images across all screening, detection, staging, prognostication, and management stages of ROP, including telemedicine applications (Table I). The review will address current issues and challenges of AI integration into clinical practice and discuss potential solutions (Fig. 1).

TABLE I
A SUMMARY OF PREVIOUS RELATED REVIEW

| Survey | Year | Screening | Detection | Staging | Prognosis | Management | Telemedicine | CBIA | ML | DL | Major Findings |
|---|---|---|---|---|---|---|---|---|---|---|---|
| Scruggs et al. [18] | 2020 | - | x | - | - | - | x | - | x | x | AI improves ROP diagnosis globally by adding objectivity and efficiency, aiding in disease screening and education, and enabling quantitative disease monitoring. |
| Gensure et al. [19] | 2020 | x | x | - | - | - | - | - | x | x | Real-world AI implementation for ROP diagnosis requires efforts in data standards, validation, and addressing ethical, technical, clinical, regulatory, and financial considerations. |
| Barrero-Castillero et al. [20] | 2020 | - | x | - | - | - | x | - | x | x | The shortage of ophthalmologists for ROP is a serious concern. AI image analysis combined with clinical information shows the potential to predict ROP risks and address the widening workforce gaps. |
| Azad et al. [21] | 2020 | x | x | - | - | x | x | - | x | x | Teleophthalmology is gaining popularity in regions with limited expertise in managing ROP. AI is useful for diagnosing, monitoring, and managing ROP, and for academic purposes. Visual rehabilitation, while often overlooked, is an important aspect of ROP management. |
| Bao et al. [22] | 2021 | - | x | x | - | - | x | x | x | x | DL can detect and predict ROP through telemedicine in rural areas, but more research is needed for clinical integration. CBIA using ML/DL has high accuracy for ROP diagnosis. Multiple-instance learning needs further investigation. |
| Campbell et al. [23] | 2021 | - | x | - | - | - | - | x | x | x | Real-world care for ROP faces challenges such as diagnostic errors and limited trained examiners. Digital fundus imaging has enabled telemedicine programs and the use of AI to enhance diagnosis objectivity. The review discusses the history, early progress, and future potential of AI in improving ROP care. |
| Ramanathan et al. [24] | 2022 | x | x | x | - | - | x | - | x | x | AI techniques showed high sensitivity and specificity, comparable to ophthalmologists, and accurately differentiated severity using automated classification scores. |
| Bai et al. [25] | 2022 | - | x | - | - | - | - | - | - | x | Twelve studies measured the performance of detecting ROP and plus disease. The average AUROC was 0.98, with high sensitivity and specificity for both ROP and plus disease. However, few studies presented externally validated results or compared performance to expert human graders. |
| Sabri et al. [1] | 2022 | x | x | - | - | - | x | x | x | x | Telemedicine and AI are being used more for ROP management. CBIA is accurate for diagnosis but needs more research for clinical use. The article covers ROP pathophysiology, classification, diagnosis, screening, treatment, and recent technological advances. |
| Shah et al. [26] | 2023 | - | x | - | - | - | x | - | x | x | AI, including DL, shows promise in detecting and predicting ROP. Its potential use in rural areas through telemedicine is being explored, but questions remain about its integration into daily clinical practice. |
| Hoyek et al. [27] | 2023 | - | x | - | - | x | - | - | x | x | The review discusses using innovative technologies, AI, and biomarkers for diagnosing and treating ROP. It covers AI-based models for detecting ROP, challenges in implementing AI in clinical practice, the use of OCT and OCTA for evaluating ROP, potential biomarkers, and the importance of integrating biomarkers and AI for early disease detection in ROP screening. |
| Our Study | 2024 | x | x | x | x | x | x | x | x | x | The review discusses how AI-based methods, telemedicine, and biomarkers can improve the screening, diagnosis, staging, and treatment of ROP. Traditional methods have been subjective, but AI can reduce subjectivity and improve patient outcomes. Challenges remain in building diverse data sets and addressing ethical considerations. |

ROP: Retinopathy of Prematurity, AI: Artificial intelligence, ML: Machine Learning, DL: Deep Learning, CBIA: Computer-Assisted Image Analysis, OCT: Optical Coherence Tomography, OCTA: Optical Coherence Tomography Angiography,

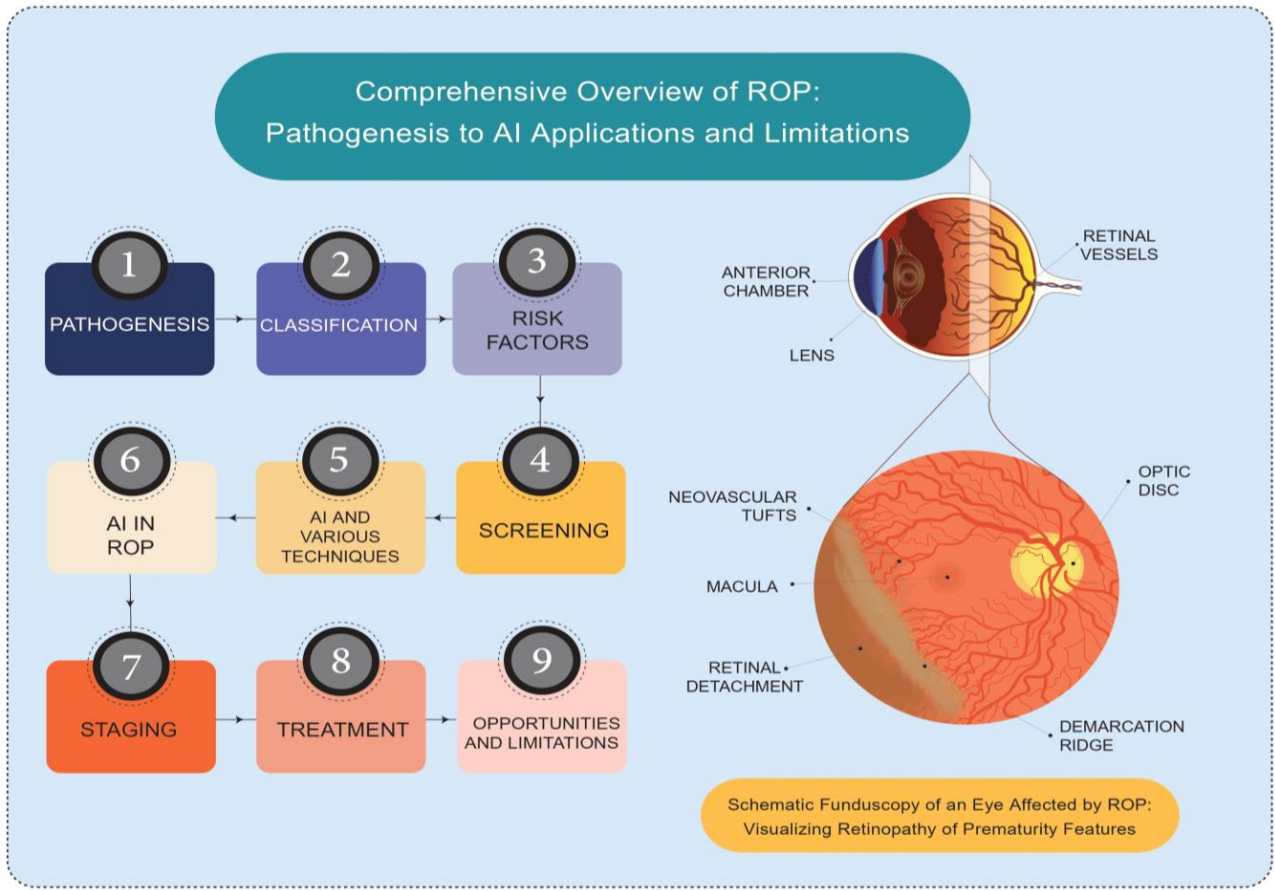

Fig. 1. Overview of AI integration into comprehensive multi-staged diagnosis and management of ROP.

## II. METHOD

We performed a comprehensive review of the literature [28] by searching across PubMed, Medline, Google Scholar, Scopus, Web of Sciences, and IEEE Xplore databases for academic papers published in the English language until December 31, 2023, on AI applications on ROP using the following MeSH keywords and their combinations: "retinopathy of prematurity," "artificial intelligence," "diagnosis," "prognosis," "therapeutics," "prematurity retinopathy," "retrolental fibroplasia," "neural network," "retinal imaging" "fundus imaging," "deep learning," "machine learning," "prediction," and "machine vision." Additional articles were obtained from the reference sections of identified publications.

## III. RESULT

The initial search yielded 1487 articles, which were then reviewed by the authors [AJ, SFM, PP]. After removing 483 duplicate articles, 883 works that did not align with our research question were excluded based on the review of the titles and abstracts. We excluded case studies, animal studies, and studies lacking clear methodology. From the full-text review of the remaining articles, we excluded another 32 articles, leaving 89 articles that met our requirements for relevance, as well as well-defined methods and technical specifications (Fig. 2). We included original studies that used artificial intelligence in the diagnosis, treatment, management, and prognosis of ROP. See the supplement for a table summarizing the included studies [29-117].

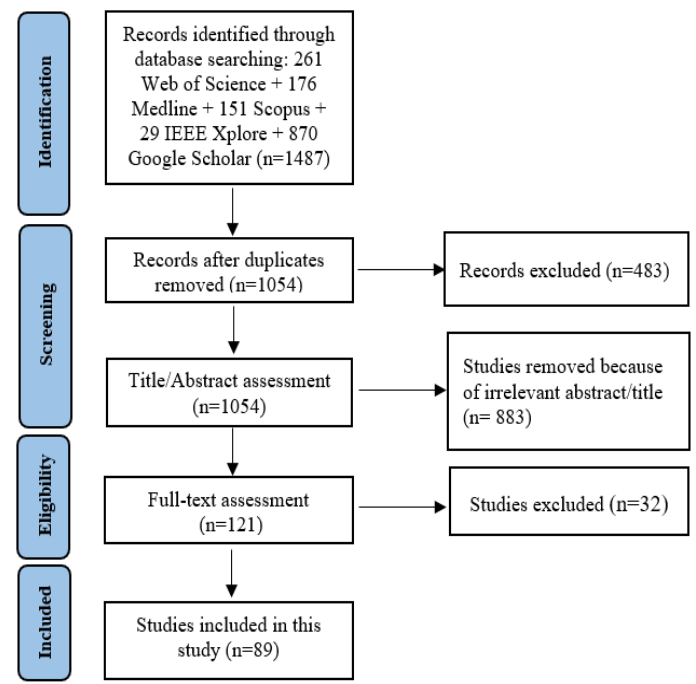

Fig. 2. Preferred Reporting Items for Systematic Reviews and Meta-Analyses (PRISMA) flow diagram for study.

## IV. DIAGNOSIS AND SCREENING OF ROP

### A. Classification of ROP

ROP was first described in the 1940s, but it was only in 1984 that the standardized International Classification for Retinopathy of Prematurity (ICROP) emerged [118]; this classification has since been updated twice, the latest in 2021 [119]. The ICROP3 recommends that the extent and severity of retinal involvement in ROP be classified separately: extent is graded based on involved zones in the retina, taking normal retinal vascular development into account; and severity, the most serious abnormality observed—vascularization abnormality (Stages 1 to 3) or retinal detachment (Stages 4 and 5)—as well as the presence of "pre-plus" or "plus" disease (i.e., tortuous arterioles and dilated venules at the posterior pole of the retina) [119].

### B. Risk factors for ROP

ROP is a disease of premature infants, and the risk factors include low gestational age, low birth weight, sepsis, intraventricular hemorrhage, blood transfusion, use of supplemental oxygen, longer duration of oxygen therapy, and mechanical ventilation [1, 120]. Hospitalized infants and those with multiple comorbidities have a higher probability of developing ROP [121].

### C. Screening for ROP and diagnosis of ROP

Infants born before 30 weeks of gestational age or with birth weights less than 1500 g are recommended to be screened for ROP at 31 weeks or four weeks of gestational and chronological ages, respectively [122], via either ophthalmologic examination or—where local expertise is limited—telemedical review of retinal images by specialists [122, 123]; and monitored, with follow-up retinal assessment schedule based on the findings as classified using ICROP3 [124]. Skilled ophthalmologists, either for examination or remote image analysis, are essential—but are often inaccessible and limited—because assessment by non-expert medical graders is significantly less accurate [125]. Even among expert ophthalmologists, considerable interobserver variability exists in identifying treatment-requiring ROP [126].

### D. Diagnosis of ROP plus disease and non-plus disease

Intervention trials of early treatment, including cryotherapy, for ROP demonstrated the significance of plus disease, characterized by arterial tortuosity and venous dilatation beyond the standard reference image, as a treatment [118]. Pre-plus was later recognized as an intermediate state characterized by aberrant vascular alterations less severe than the reference norm of plus disease [127]. Currently, the binary classification of ROP patients into plus and non-plus diseases is the most crucial step that informs decisions on laser therapy, which is indicated in the first group [128],[129]. However, diagnostic agreement on plus disease by expert interpretation—the reference standard—is imperfect, which has significant implications for the validity and robustness of clinical, telemedicine, and computer algorithm-based diagnosis of plus and non-plus disease in ROP [130-132].

### E. How to detect ROP and confirm the diagnosis?

Imaging systems have largely superseded binocular indirect ophthalmoscopic examination for ROP screening. RetCam (Clarity Medical Systems, Pleasanton, CA)— the most established—possesses advantages of user-friendliness, broad 180-degree field-of-view, fine 640 × 480 image resolution, image archival capability, and capability for real-time comparison of individual patients' progress images. RetCam-enabled digital image diagnosis of ROP is feasible and accurate [12] and has been used extensively in plus disease detection [131-133]. Beyond its clinical utility, the use of Retcam has also facilitated clinical audits, research, and teaching. NM-200D (NIDEK Inc, Fremont, CA), a non-contact, narrow-angle digital camera with a more limited 30-degree field-of-view [14], has been used for ROP screening and plus disease detection [31, 134, 135]. Similarly, the Kowa portable fundus camera (Kowa Optimed Inc, Torrance, CA) has also been deployed for image-based plus disease detection [29]. Smartphone-based fundus camera technology has been proposed as a low-cost, population-based patient- and physician-facing screening and monitoring tool for central and peripheral retinal diseases [136]. While this approach has the advantages of fine image resolution, accessibility, portability, and cost-effectiveness, the smartphone field-of-view is currently only 65 degrees, which limits the capture and assessment of the periphery of the retina and determines the ROP zone or stage [13]. Nevertheless, it has been shown feasible to have health workers trained to use smartphone equipment to collect retinal images for remote analysis by ophthalmologists [137]. Telemedicine screening of digital retinal images, whether acquired in the clinic or community, promises a solution for low-resource countries that can potentially minimize the health and societal impacts of ROP [138, 139].

### F. Challenges in ROP screening and diagnosis

Delayed detection and treatment of ROP result in macular fold formation, retinal detachment, and blindness by disrupting the normal development of retinal blood vessels [140]. Universal screening of premature newborns carries a high-cost burden. In the United Kingdom, 55 at-risk infants are evaluated for each baby treated [141]. Considerable interobserver variability and treatment variations contribute to the complexity of weighing efficacy benefits versus the cost of ROP screening [142, 143]. Compounding these difficulties, there is a shortage of qualified workforce and well-trained ophthalmologists to perform and interpret retinal scans, respectively. In resource-limited countries, challenges arise from constrained healthcare infrastructure, a lack of diagnostic imaging equipment, and a shortage of skilled personnel. These real-world constraints contribute to difficulties in achieving accurate and timely ROP diagnosis. Consequently, the full potential of ROP screening in preventing vision loss is compromised [32], particularly when patients present late with already poor vision prognoses or severe retinal detachments [140]. Addressing these challenges demands focused efforts to improve access to diagnostic tools as well as to train, and standardize the training of, more skilled medical personnel and ophthalmologists. Regarding the latter, few centers in Asia, including middle-income countries, offer surgery as a treatment option for advanced ROP, which underscores the scarcity of experienced pediatric retinal surgeons and pediatric anesthetists in this part of the world.

### G. Telemedicine for ROP

Tele-retinal image analysis by experts is accurate and feasible and is particularly useful in hospitals and clinics with limited access to skilled examiners [144]. Challenges to remote image interpretation include image quality issues, confounding by background fundus pigmentation, geographic variations in disease phenotype, and interobserver variability of expert interpretation [145].

### V. An all-encompassing solution; AI

The widespread adoption of fundus photography for the detection and follow-up of diverse retinal diseases has shifted the diagnostic paradigm from clinical eye examination to offline retinal analysis. Automatic computer-assisted diagnostic (CAD) technologies can aid ophthalmologists in making a risk-free, precise, and economical diagnosis of ROP. Computer-based image analysis (CBIA) has mitigated inherent human biases [146], as well as reduced the cost and time burdens, of manual expert interpretation [147]. Analogous to significant inroads made in AI-assisted management of other eye diseases [148], AI promises to overcome the challenges of retinal image analysis for ROP detection arising from inconsistencies in retina image quality, variations in patient phenotypes, and variabilities of expert interpretations, potentially leading to improved diagnostic accuracy, reduced burden on healthcare professionals, and better clinical outcomes [149]. The successful implementation of AI integration into CBIA of ROP depends on consistent standards for data collecting, rigorous external validation of the candidate AI model, and demonstration of its feasibility in real-world deployment AI. Indeed, several AI algorithms have been developed, and validated and are clinically adopted for the screening of ROP and follow-up of patients who require treatment [150].

### A. AI methods in ROP

*Early AI systems for ROP detection (Semi-automated):* CBIA introduces objectivity into the diagnosis of plus disease, which can otherwise be subjective, qualitative, and variable with human experts [19]. Early ROP diagnostic systems—ROPTool, RISA, VesselMap, and CAIAR—were designed to extract specific vessel features from retinal images, on which either manual or semi-automated methods were applied to derive quantitative measures of vessel tortuosity and dilatation to determine the presence of plus disease in ROP..

A software application called ROPtool, which Duke University developed, was designed to extract tubular structures from multi-view images, analogous to techniques used to analyze intracerebral blood vessels on magnetic resonance angiography [151]. ROPtool attained an excellent 97% sensitivity for classifying patients into plus, pre-plus, vs. non-plus categories after being trained on 185 RetCam fundus photographs labeled by six pediatric ophthalmologists [152, 153]. In modifications to ROPtool, Wallace et al. developed methods to compute both vessel tortuosity (dividing the length of a smooth curvature fitted along points on a venule or arteriole by the total length of the vessel in a 30-degree region placed on the optic nerve) and diameter (measured from cross-sectional profiles of the vessel, and normalized to the distance between the optic disc and the middle of the macula to account for variations in image magnification [154]). The model attained areas under the receiver operating characteristic curves (AUC ROCs) of 0.93 and 0.90 for identifying dilatation sufficient for plus disease and pre-plus disease. The sensitivity and specificity for identifying dilatation consistent with plus disease were 89% and 83%, respectively [36].

Imperial College London pioneered research on the use of semi-automated analysis of RetCam images to identify and quantify retinal vascular abnormalities, which led to Swanson et al. developing the Retinal Image Multi-Scale Analysis (RISA) algorithm [155] that was later modified by Martinez-Perez et al. [156]. To measure tortuosity and dilation of arterioles and veins in the retina, RISA incorporates a multiple-pass region-growing technique to segment vessels based on feature information of the eight neighboring pixels surrounding each pixel. Geometric elements such as—but not limited to—highest gradient, primary curvature, and first and second derivatives of image intensity are used. RISA first locates the maximum scale space to expand an area, then adds a restriction to constrain the gradient [157]. Compared with manual contouring of vessel borders by human experts—the reference standard for ROP diagnosis [158]— RISA was accurate for diagnosing plus disease [33, 158, 159]. Additionally, vessel dilation can be quantified by computing the mean vessel diameter (in pixels) as the ratio of the entire area of the vessel to its length [156]; and tortuosity, by two indicators: combined curvature (radians/pixel, Integrated curvature), the totality of all angles along the blood vessel normalized to the vessel length; and tortuosity index, the ratio of vessel arc length and the linear distance between its starting point and endpoint (Fig. 3). Compared to manual segmentation by experts, RISA was shown to have a specificity and sensitivity of 64% for the venous tortuosity index, 70% for the arterial tortuosity index, 60% for analysis of the arteriolar diameter, and 76% for venular diameter. When quantitative assessment of vascular dilation and tortuosity was included, both sensitivity and specificity rose to 94% [33].

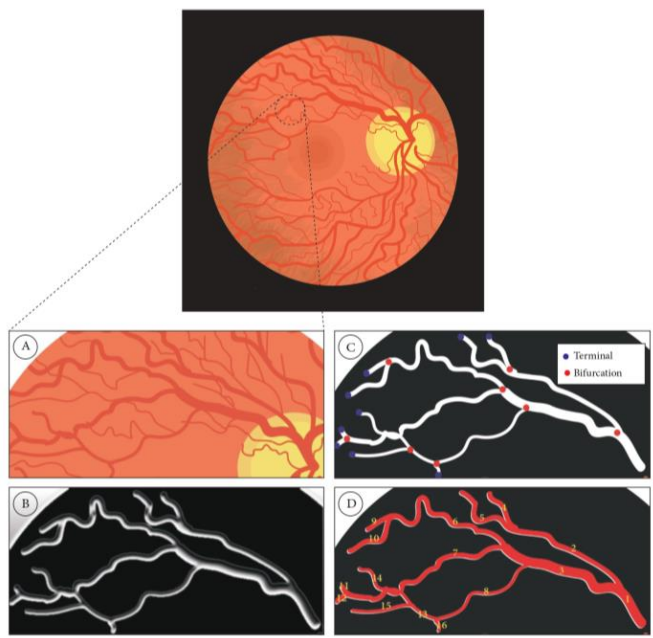

Fig. 3. Demonstration of Retinal Image Multi-Scale Analysis (RISA). (A) Cropped retinal image, (B) selected vessel. (C) vessel is segmented, (D) the skeleton is constructed, and (E) the vessel is tracked. After these steps, the program measures dilation, tortuosity index, and integrated curvature.

Rabinowitz et al. introduced VesselMap, a semi-automated software that calculates the diameters of retinal venules and arterioles by analyzing brightness indices perpendicular to the lengths of vessels found within two-disc diameters of the optic nerve head. The study aimed to clarify whether early retinal vascular measurements could predict the development of severe ROP in high-risk preterm infants [31]. Seventy-eight retinal images of at-risk newborns at 31 to 34 weeks gestational age were captured using a non-contact camera and stratified into three groups based on disease and its progression: no ROP, less severe ROP that did not require treatment and severe ROP that later progressed to require treatment. VesselMap was used to measure the diameters of major superior and inferior temporal arteries and veins, which were compared among the groups. The mean diameters of all four main temporal vessels of the retina were significantly larger in eyes that progressed to severe ROP requiring treatment after adjusting for birth weight, gestational, and chronological age. Early evaluation of retinal vessel diameter demonstrated predicted severe ROP with AUC scores ranging from 0.75 to 0.94.

Computer-aided image analysis of the retina (CAIAR), conceived as an enhancement of the RISA software, uses a high-probability scale-space comprising filters sensitive to ridge-like patterns across four distinct scales for extracting vascular structure from retinal images, followed by comparison of the best-fitting vessels to the program output [34]. Gaussian profile analysis of vessel dimensions (diameter, length, and orientation) is then performed, and the Gaussian filter's second derivative is used to determine whether the middle of the vessel is the region with the highest contrast. Fourteen methods were evaluated to assess tortuosity and two to calculate vessel width on 60-degree images. While the semi-automated tortuosity measures matched well with human expert assessment, the vessel width measures did not match.

These early ROP diagnostic systems share a similar design:

1. Extract vessel features from retinal images.

2. Perform vessel segmentation.

3. Measure vessel tortuosity and diameter using manual and/or semi-automated methods.

A limitation involves inaccurate vessel segmentation, which may require manual vessel delineation (especially in the posterior pole) or morphological pre-processing before semi-automated vessel segmentation [160]. Mathematical methods have been proposed for measuring vessel tortuosity—such as length-to-chord [30, 161], angle-based [162], curvature-based [163, 164], or spatial frequency measures [29]—but did not always correlate with expert clinical assessment of tortuosity [165]. The performances of these systems exhibit varying degrees of diagnostic correlation with clinically ascertained ROP [37].

*Optical coherence tomography and AI*: Optical coherence tomography (OCT) uses the interference of wavelengths of light reflected off superficial layers of a tissue sample, such as the retina, to generate three-dimensional images with micrometer axial resolution. It is widely used to diagnose retinal disorders, including ROP [119]. Using a portable device version of advanced spectral domain OCT to image the retina of newborns with ROP, Chen et al. were able to obtain images of the affected retinal microstructures at the vascular-avascular junction with histology-like detail [166]. Due to the limitations of the OCT angle, investigations have traditionally focused on abnormalities in the posterior part of the retina. OCT devices with a wider field of view and faster image acquisition speed are being designed to maximize the positive effects of OCT on ROP [166-168]. Using a handheld OCT device with a wider 55-degree angle that depresses the sclera simultaneously, Scruggs et al. showed that it was feasible to visualize and assess the periphery of the retina to evaluate ROP disease [83].

For early prediction of referral-warranted ROP in at-risk infants, Legocki et al. developed and compared multivariate models incorporating clinical data and imaging data: key variables included gestational age, birth weight, as well as non-contact handheld spectral domain OCT-assessed vessel elevation, hyporeflective vessels, and vitreous opacity ratio. Simple models that included either only clinical variables (gestational age and birth weight) or imaging markers attained AUCs of 0.68 (77.3% sensitivity, 63.4% specificity) and 0.88 (81.8% sensitivity, 84.8% specificity), respectively. A generalized linear mixed model and an ML model that analyzed clinical and imaging data attained AUCs of 0.94 (95.5% sensitivity, 80.7% specificity) and 0.83 (91.7% sensitivity, 77.8% specificity), respectively. These observations support the relative and incremental predictive utility of OCT while not confirming the utility of the deployed ML model [102].

### B. ML and DL models

Unlike ML and DL-based solutions, the aforementioned fundus photograph-based and OCT-based CBIA systems for ROP assessment rely on at least some degree of manual

input—in semi-automatic models, humans will still need to identify or choose specific findings within the images—to generate quantitative measures for plus disease assessment. Overall, the correlations with clinical diagnoses were modest. While these systems may not represent true learning in the traditional sense, they are nevertheless the foundation for the development of modern ML and DL-based systems [38].

*ML methods in ROP:* Earlier non-ML systems have modest performance for ROP diagnosis. For example, ROPTool™ showed 96% sensitivity and 64% specificity for the detection of plus versus non-plus (without pre-plus disease identification) and 91% sensitivity and 83% specificity for any vascular anomaly (1 or more on a scale of 0-4) [44]. Many researchers worked to improve upon this by incorporating ML. In an ML model, features related to the disease are defined (dilation, tortuosity, ridge presence, etc.) in the input signal (retinal image) and mathematically transformed into usable forms using explicit algorithms (feature extraction). The data are then divided into training and testing sets. A classifier is chosen to train the model to maximize associations between the features and established ground truth labels. The learning is then used to evaluate model performance on the test dataset.

Given that the model elements are manually crafted, their optimality is not assured. Potential issues include features that may not consistently align with the ground truth. These feature extraction methods might inaccurately quantify the features, classifiers that could inadequately learn or overfit training data, and ground truth labels that may exhibit noise or lack generalizability. The presence of these factors can adversely impact model performance [19].

Capowski et al. [29] devised a quantitative index to assess the progress in tortuosity of retinal vessels over one to two months, which, when fed to a computer program, could discriminate structural modifications found in ROP vs. non-ROP conditions. In addition, they observed that plus disease-induced arteries become tortuous with a distinctive spatial frequency, which forms the basis for a diagnostic numerical indicator. Heneghan et al. [30] developed a software called Vessel Finder, which assessed the tortuosity and dilation of segments of vessels using a binary mask. The segmentation followed a multi-stage process involving morphological pre-processing to highlight linear vessel-like features; application of morphological filter based on a second derivative operator to enhance vessel visibility; thresholding to generate a vascular mask (careful tuning of threshold settings is essential to optimal segmentation); skeletonizing the mask to locate vessel bifurcations and crossings, in order that the width and tortuosity of vessel segments can be delineated and quantified. The narrowest part of the vessel at each location was determined, and the average width for the entire vessel segment was computed. Tortuosity was defined as the ratio of vessel length to the shortest straight-line distance between its two ends. On test photograph, the model width and tortuosity assessments were demonstrably accurate, and on a sample of 23 participants (eleven with ROP and needed treatment; nine had no ROP; and three had spontaneous ROP regression), the model attained sensitivity and specificity of 82% and 75%, respectively, based on vessel width and tortuosity measures.

Cansizoglu et al. developed the i-ROP method, which classified fundus photographs into predicted plus, pre-plus, and normal classes using either manually cropped nested circular and rectangular retinal images of varying sizes centered on the optic disc or eleven extracted quantitative vessel dilatation and tortuosity measures extracted from pre-processed vascular trees constructed from vessel centerlines [40]. The model trained on a dataset of 77 wide-angle retinal images (14 plus, 16 pre-plus, and 47 normal) of at-risk newborns that had been labeled by three experts based on an ophthalmoscopic examination and attained 95% three-class classification accuracy that was similar to expert diagnosis, and superior to the mean performance of 31 non-experts. Notably, a cropped circle of 6 optic disc diameter (which extends beyond the posterior pole of the eye) and point-based (as opposed to segment-based) measurements of combined arterial and venous tortuosity yielded the best discrimination. Campbell et al. [45] compared i-ROP vs. manual interpretation by eleven experts on a dataset of 73 retinal images. The results confirmed similar 95% accuracy and 87% mean accuracy (range 79%-99%), respectively. Notably, the i-ROP system neither misclassified normal eyes as plus disease eyes nor any plus disease as normal. It was also observed that the experts tended to deviate from the defined ROP diagnostics standards that confined the analysis to the posterior pole (which is about 2 to 3 optic disc diameters in size) and considered the tortuosity of arteries and veins beyond this region. This observation is corroborated by prior literature. Rao et al. showed that the agreement among 13 experts for diagnosis of plus disease in wide-angle images was superior to medium-angle and narrow-angle images, suggesting that physicians routinely analyze peripheral and mid-peripheral zone data in a manner that is not encoded in the definition of plus disease [169].

Similarly, in another study employing qualitative research methodologies to investigate experts' diagnostic processes [170], peripheral retinal vascular characteristics were highlighted as helpful for the plus disease diagnosis. The tortuosity of veins, rather than arteries, was higher in the standard published image [171]. These findings align with the outcomes of Wilson et al., who clarified that the quantitative tortuosity of the arterial and venous systems increased with the increasing clinical stage of ROP. These findings remained accurate regardless of whether four or eight arteries were examined[38]. From the model development perspective, this is a valuable input as it obviates the need for the computer system to differentiate between retinal arteries and veins (at least in the assessment of tortuosity), which can be challenging, even for human experts [152]. In contrast, quantitative analysis of vascular dilation, influenced by external factors like image magnification, is of lesser diagnostic significance in ROP, which aligns with other ROP algorithms [44]. A limitation of early i-ROP systems is the need to manually segment the retinal vessels on the image, although fully automated systems to address this limitation have been developed [45].

Apart from i-ROP, some researchers adopted different approaches to enhance the sensitivity of semi-automatic

algorithms—e.g., ROPtool, RISA, etc.—that assess the tortuosity and dilatation of vessels in retinal images [38]. Pour et al. [47] developed software incorporating a range of ML classifiers to classify non-plus ROP, pre-plus ROP, and plus ROP classes. On 87 retinal images, the model attained 72.3%, 83.7%, and 84.4% accuracy for diagnosis of the plus, pre-plus, and non-plus images, respectively.

*DL in ROP:* DL models are end-to-end systems that do not require data processing, including segmentation and feature extraction, and are therefore preferred over ML methods [172]. For instance, early ML-based i-ROP models can discriminate plus, pre-plus, and non-plus diseases. However, retinal blood vessels must be manually segmented before input into the system. Not surprisingly, newer systems now mostly use convolutional neural network (CNN)-based DL methods. Indeed, DL algorithms have demonstrated improved accuracy in the detection of plus and pre-plus diseases [16]. Campbell et al. [88] introduced i-ROP-DL, which could automatically generate ophthalmoscopic staging and vascular severity scores directly from inputs of retinal images. These scores performed as accurately as expert committee members, who graded each image with plus value and stage ranging from 1 to 9 and 1 to 3, respectively.

CNNs, widely used in medical imaging analysis [173, 174], learn to prioritize image characteristics that best relate the input image with the diagnostic label by training on large datasets and adapting learnable weights and biases within the network architecture. This task is accomplished without direct human input, with or without pre-processing [50, 175]. Worrall et al. [176] were the first to study a fully automated DL-based ROP detection model. First, a pre-trained GoogLeNet model was adjusted to serve as an ROP detector: The researchers adjusted a pre-trained GoogLeNet to function as a detector for ROP. By making minor modifications, they produced an estimated Bayesian posterior, expressing the probability of ROP presence based on observed data and model adjustments. Second, they trained a second CNN to provide innovative feature map representations of diseases learned directly from the data, significantly facilitating the grading process. The system's image detection classifier was almost as accurate (92%) as human experts.

The low contrast of newborn retinal images makes it challenging to distinguish between the demarcation line that divides the peripheral and vascularized retina and the appearance of a ridge with a width, which characterizes Stage 2 ROP. Mulay et al. [54] developed a regional CNN-based algorithm, Mask R-CNN, that employs image pre-processing via normalization to compensate for low-quality input to facilitate the identification of ridges. The model was trained on 175 of the 220 labeled retinal images of 45 newborns in the KIDROP project to segment the ridge area to the ground truth, and tested on the remaining 45 images. The model attained 88% detection accuracy for early-stage ROP.

Recently, Attallah et al. [78] introduced DIAROP. This novel ROP diagnostic model uses four pre-trained CNNs to extract spatial features by transfer learning. It then reduces the dimension of spatial features extracted in the preceding phase and integrates them using the Fast Walsh Hadamard Transform. The model attained a high 93.2% accuracy and 0.98 AUC score. Salih et al. [100] also applied transfer learning using 1365 retinal images to train four CNN-based networks—VGG-19, ResNet-50, and EfficientNetB5—to recognize ROP zones. EfficientNetB5 demonstrated a superior accuracy of 87.27% vs. the other three architectures, while the voting classifier attained an aggregate accuracy of 88.82%.

Other CNN networks have also been used for ROP diagnosis. Region-based CNN (R-CNN) encompasses two phases: the first involves selecting candidate areas in an image that may or may not include a particular object; the second is the regional classification of the object. Deep neural networks (DNNs) are artificial neural networks (ANNs)—which always contain elements of neurons, synapses, weights, biases, and functions [177]— designed to mimic how the human brain learns with multiple layers of complexities (i.e., mathematical manipulations [178]) between the input and output layers [179]. For example, a DNN analyzes the input image and calculates probabilities for various categories in an image classification task. Following evaluating these probabilities, the user can set criteria, such as a probability threshold, to filter and determine the suggested label for the identified category. Kumar et al. [98] combined a deep CNN with image processing to automatically identify retinal attributes of retinal blood vessels and optical discs, as well as ROP disease classification, using a rule-based approach. YOLO-v5 was employed for optic disc detection, and Pix2Pix, U-Net, or another deep CNN was employed for blood vessel segmentation. Training was performed on public retinal image datasets (1,117 for Optic Disc; 288, blood vessels), and testing was performed on 439 preterm neonatal retinal images to evaluate the presence and severity of ROP in different retinal zones (Zone I, Zone II, and Zone III). Additionally, six blood vessel masks— the retinal image regions corresponding to blood vessels— were assessed to analyze abnormalities in blood vessel development within the retina. The approach yielded excellent results:

- 98.94% accuracy (when Intersection over Union 0.5) for optic disc detection
- 96.69% accuracy (Dice coefficient 0.60 to 0.64) for blood vessel segmentation
- 88.23% accuracy for Zone-1 ROP diagnosis

Jemshi et al. [101] prioritized sensitivity for detecting plus and non-plus disease in their efficient artificial neural network architecture incorporating wavelet and curvelet transforms of retinal images to derive additional features of transform energy coefficients alongside vascular features. On a 178-image dataset comprising 81 plus and 97 non-plus cases, the model attained excellent 96% accuracy, 93% specificity, and 100% sensitivity, which was the prime objective.

Wang et al. [52] created DeepROP, which comprises two DNNs with distinct functions: Id-Net for detecting the presence or absence of ROP characteristics; and Gr-Net, for grading ROP instances as mild, moderate, or severe. DeepROP was trained on a large dataset of retinal images encompassing a wide range of ROP features and various degrees of severity and attained 84.91% sensitivity and 96.90% specificity in ROP diagnosis.

While the results suggest that ROP detection will be successful in the future, ROP grading was not. These results could be attributed to the less clear demarcation between "normal" and "ROP" instances compared to the differentiation of "minor" and "severe" ROP cases. Moreover, Gr-Net used more limited image datasets compared with Id-Net — there were also few "severe ROP" instances among the training data—which could explain the relatively poorer performance.

The i-ROP-DL system, developed for plus disease classification by the Imaging and Informatics in ROP (i-ROP) consortium [50], encompasses a two-CNN framework: U-Net for vessel segmentation and Inception-V1 for classification. A CNN was trained on 5,511 retinal images, which had each been independently graded by three experts, as well as labeled by one expert into normal, pre-plus disease, or plus disease (the reference standard diagnosis). The algorithm was tested on 100 photos outside the training set and assessed using 5-fold cross-validation. The i-ROP cohort research gathered images from 8 different universities. Eight ROP specialists were used to evaluate the performance of the DL method. The algorithm achieved an AUC-ROC of 0.94 for normal retina diagnosis and 0.99 for diagnosis of plus disease. Additionally, the algorithm attained an accuracy of 91% [50]. Inconsistencies in plus disease diagnosis led to clinically significant disparities in medical management [126], even though evidence-based ROP management guidelines rely on treating patients according to the existence of plus disease to avoid vision loss and blindness [128, 180]. For a data set of 34 images, Chiang et al. in 2007 studied the diagnostic accuracy of 22 ophthalmology specialists for plus disease. Only four images had a complete consensus for plus disease [131]. Multiple articles since then have revealed similar findings, showing that expert pairs had fair (0.21-0.40) [181, 182] to moderate (0.41-0.60) [131, 138] agreement in detecting plus disease. Due to systematic bias in plus disease diagnosis, an inherent limitation in ROP research studies, it was uncertain whether these disparities extended to practical variations in treatment or outcomes.

Pre-trained on the publicly accessible ImageNet database, DeepROP and i-ROP-DL methods exhibit excellent performance and good agreement with expert opinion. Recent research has shown that an ROP severity score obtained from DL may be used for disease screening through the i-ROP-DL classifier [183]. The ROP vascular severity score helps evaluate the rate of disease development [184], tracks therapy effectiveness [185], and discriminates between aggressive-posterior and less severe forms of ROP [64].

Recently, there have been numerous reports about automated systems. There are two types of automated systems: those that rely on CNNs for feature extraction and those that rely on systems that extract characteristics manually (such as vascular tortuosity and vessel diameter). However, Brown et al. [50] used CNNs for segmenting retinal arteries and veins, detecting additional pathology like plus disease, and following patients receiving therapy for ROP. When comparing normal to pre-plus or worse and to not-plus illness, their CNN method obtains AUC scores of 0.94 and 0.98, respectively. In addition, Worrall et al. [186] suggested a CNN model for identifying plus disease that obtained 92% accuracy. In many medical imaging applications, CNNs have been reported to have enhanced performance compared with feature-extraction-based ML approaches [187]; nevertheless, they have the drawback that the CNN features are not visible or explainable.

Yildiz et al. explored the possibility that combining a CNN model for identifying the relevant vascular structures and a previously developed feature-extraction algorithm could result in an automated plus disease classifier with performance like CNNs but with explainable features [188]. Cansizoglu presented a method to assign a severity value to a retinal picture; this study inspired the I-ROP ASSIST system [39]. While the system introduced in Yildiz's investigation used a similar approach to vessel tracing and feature extraction, there are significant differences. Rather than relying on the vessel segmentation approach previously used, they completely automated the system by adding a detector to the center of the optic disc [188].

The i-ROP consortium showed that a hybrid system, called i-ROP ASSIST, could attain CNN-like performance. It achieved this by employing the effective U-Net-based segmentation algorithm for improved vessel delineation and then using a conventional feature extraction and classification approach, allowing for explainable features. In contrast, the investigations utilized Gabor filters to categorize 110 images of 41 individuals into ROP disease plus/non-plus categories [189, 190]. Besides that, another study used 20 fundus images to develop an ROP tool that could differentiate between pre-plus and plus ROP problems [191].

Similar to the diagnostic approach presented in the study by Campbell et al., "i-ROP" used an SVM classifier with a 95% accuracy rate (with 77 images) to distinguish between pre-plus, healthy, and plus ROP disease. In 2015, Cansizoglu et al. presented an ML model designed to automate the diagnosis of plus disease. This model demonstrated comparable performance to human experts. In contrast to earlier systems, this model incorporated conventional features but used an SVM to identify the most practical combination of features and field of view that exhibited the strongest correlation with expert diagnosis. An SVM is a supervised ML classifier to discern the connection between features and diagnostic outcomes. The system achieved its highest accuracy (95%) when it combined vascular tortuosity measurements from arteries and veins with the widest field-of-view (i.e., a radius of 6-disc diameters) [40].

Interestingly, when the images were trimmed to match the field of view of a standard photograph, the accuracy dropped to less than 85%. This finding suggests that clinicians consider vascular information from a larger retinal area than what is depicted in the standard photograph. Although this system demonstrated expert-level performance, its clinical utility was limited due to its requirement for manual tracing and segmentation of the vessels as inputs [40].

Early classification models were developed using a small number of low-resolution images. These methods rely on human operators to extract vessel characteristics and segment the vessels, which might introduce error and bias into the diagnostic process when choosing the target vessels.

Segmentation and feature extraction are only two examples of time-consuming image processing techniques. We need more automated, reliable technologies, such as those based on DL models, to address this. There have been many new releases of DL-based CAD solutions for ROP diagnosis. The tools relied on transfer learning [192], which involves reusing CNNs learned on one classification challenge with massive data sets, such as ImageNet, on a separate classification challenge with fewer pictures, such as the one at hand. Transfer learning has shown the capacity to enhance diagnosis accuracy across various medical specialties[193, 194]. As a result, it is used to develop CAD instruments for ROP diagnosis.

In another paper, an innovative attention-aware and deep supervision-based network (ADS-Net) for ROP detection (Normal or ROP) and ROP grading (three levels) was developed, which achieved 0.9552 and 0.9037 in ROP screening and grading for Kappa index [84].

Lin et al. explored the correlation between time-series oxygen data from electronic health records (EHRs) and the emergence of type 2 or TR-ROP. Data from 230 infants, including demographics, GA, birth weight, and oxygen metrics, were used to develop ML models. Notably, combining demographics and oxygen data, the multimodal long short-term memory (LSTM) model outperformed the best ML models and GA-trained support vector machine (SVM) models, achieving a mean AUC of 0.89±0.04. This approach showed promise for early identification of severe ROP risk and advancing our understanding of the condition's progression [99]

*xAI in ROP:* Explainable Artificial Intelligence (xAI) is a growing interdisciplinary field focused on making artificial systems understandable to humans[195]. It explores methods to clarify complex artificial systems, which is crucial when they lack transparency[196]. xAI aims to meet the needs and expectations of human stakeholders, such as users, developers, and regulators, by enhancing the comprehensibility of these systems. These stakeholder interests are called "stakeholders' desiderata"[197].

### C. Staging of ROP

International Classification of ROP explained the staging system of ROP in 5 stages, including:

1. Zone Classification: This system defines three retinal zones based on vascularization and introduces the term "notch" for ROP lesions extending into a more posterior zone.

2. Plus and Pre-plus Disease: Plus disease involves prominent vessel dilation and tortuosity, while pre-plus disease represents less severe vascular abnormalities within Zone I.

3. Stage of Acute Disease (Stages 1–3): Stages 1-3 describe ROP's progression at the vascular-avascular juncture, with the most severe stage determining classification if multiple stages are present.

4. Aggressive ROP: "Aggressive ROP" replaces "aggressive-posterior ROP," acknowledging a severe, rapidly progressing form affecting various zones.

5. Retinal Detachment (Stages 4 and 5): Retinal detachment is divided into stage 4 (partial) and stage 5 (total). Stage 5 has subcategories based on optic disc visibility and funnel configuration.

This system guides clinical assessment, ensuring timely intervention for affected premature infants [198]. These five stages have been visually represented in Fig. 4.

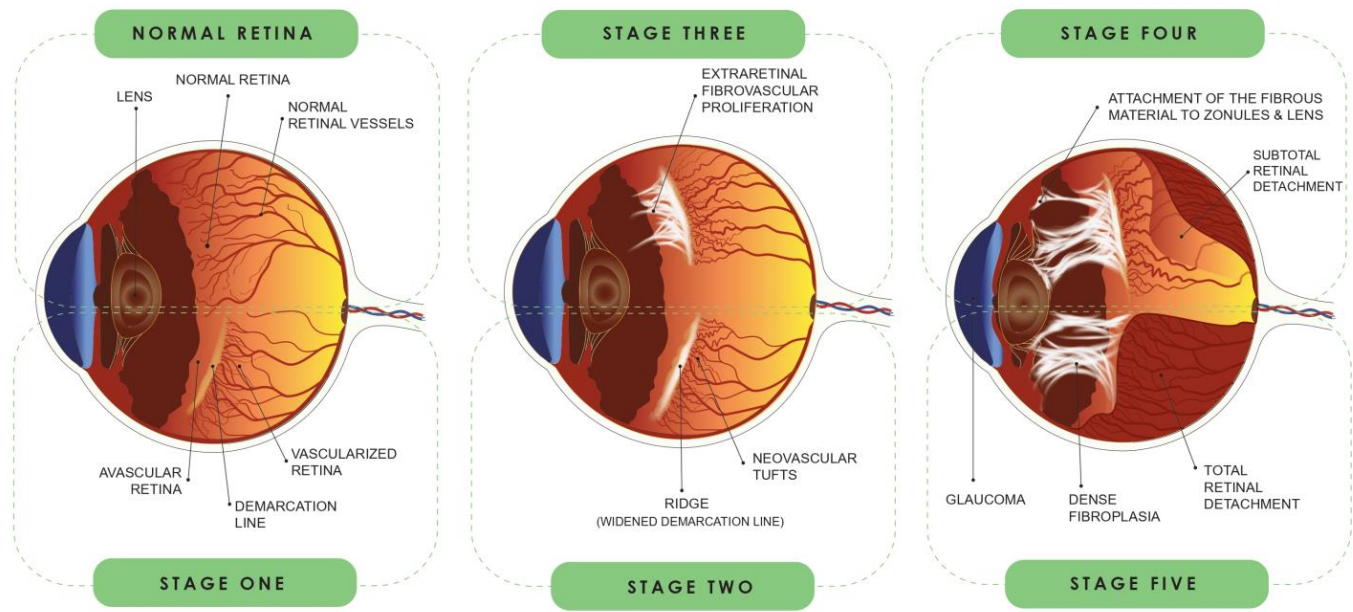

Fig. 4. Illustration of the sequential stages of ROP pathogenesis.

Unlike screening and detecting plus disease using AI in ROP, staging and scoring the severity of retinal involvement has not been as thoroughly studied [1, 15]. The first study to aid in assessing disease severity via AI was The ImageNet pre-trained Google Net that only focused on detecting plus disease [43]. While this is a standard system in diagnosis, staging, classification, and treatment of ROP is lacking, and a notable difference of opinion among individual experts and ophthalmologists exists in each of these aspects [142, 199, 200].

In a study that used a 4-stage system (normal, mild, semi-urgent, urgent) and a staging system to classify ROP fundal images via DL clinically, two CNNs were used: the 101-layer ResNet (classification network) and the Faster R-CNN (identification network). The results showed that the system could achieve an accuracy of 0.903, a sensitivity of 0.778 with a specificity of 0.932, and an F1-score of 0.761 for grading the ROP cases as "normal," "mild," "semi-urgent," and "urgent." The outcomes indicated that the system could accurately distinguish between the four-degree classifications with respective accuracies of 0.883, 0.900, 0.957, and 0.870. Misclassification was also more common among human experts than the DL system [68].

On the other hand, Peng et al. used an automated staging system to assess the severity of ROP and group each patient into one of the five groups already recognized by the ICOP3. This study used a three-stream parallel framework including ResNet18, DenseNet121, and EfficientNetB2 as the data extractor. The features from these streams were fused deeply to generate a more accurate and comprehensive feature. An ordinal classification strategy was used to improve the staging performance of the algorithm. The ROP staging network underwent assessment using both per-examination and per-image approaches. An assessment of 635 retinal fundus images from 196 examinations for per-image ROP staging was conducted. The results demonstrated an impressive performance with figures of 0.9055 for weighted recall, 0.9092 for weighted precision, 0.9043 for weighted F1 score, 0.9827 for accuracy with 1 (ACC1), and 0.9786 for Kappa. Per-examination ROP staging was carried out to assess the proposed method further using 1173 examinations and a 4-fold cross-validation strategy. This approach aimed to confirm the validity and advantages of the proposed method [74].

Similarly, in an investigation that used the images from the ICROP3 preparation process, 34-panel experts were initially asked to stage standard and typical retinal images into stages 1 through 3 and the presence of plus disease with severity from 1 to 9. The mean weighted kappa and concordance correlation (CC) for all interobserver pairs for plus disease image comparison were 0.67 and 0.88, respectively [201]. Later, the images were labeled with a DL-derived score by the iROP-DL algorithm for the severity of vascular disease from 1 to 9. The vascular severity score exhibited a significant association with both the mode plus label (P<0.001) and the ophthalmoscopic diagnosis of the stage in the same eyes (P<0.001). Additionally, it demonstrated a strong correlation with the average disease severity, with a correlation coefficient of 0.90 [201].

While these results are promising in transitioning from a qualitative to quantitative assessment of disease severity, interobserver variability remains a limitation in achieving a single standard scale for the classification of disease severity in ROP in particular. The summary of implemented AI models in ROP is visualized in Fig. 5.

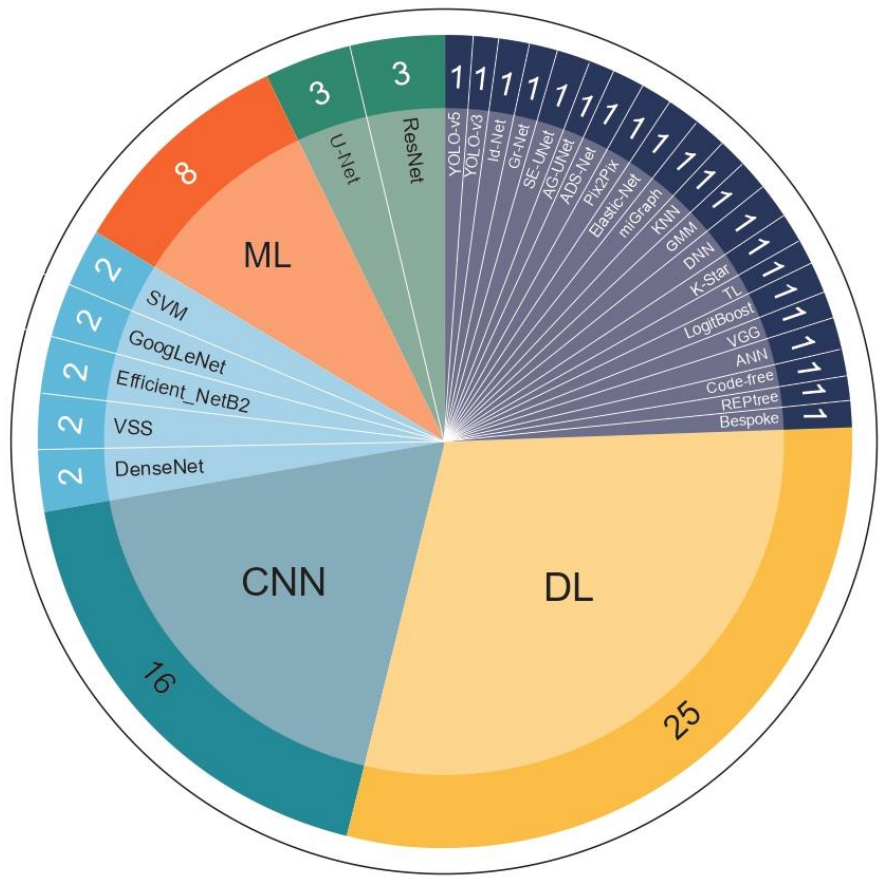

Fig. 5. Pie chart illustrating the AI methods employed for ROP detection.

### D. Treatment and prognostication of ROP

Currently, the standard treatment for the early stages of ROP includes laser therapy, with surgery being reserved for more aggressive cases [202]. Anti-VEGF therapy has also emerged as a potential adjunct or replacement for the treatment of ROP; however, the evidence for and against its implementation remains inconclusive thus far [203]. The presence of plus disease, zone I involvement, higher severity of stage, higher circumferential involvement, and rapid progression of ROP have been associated with a less favorable macular outcome after therapy [204, 205]. Rescue treatment for ROP is common in patients with severe refractive errors and those requiring spectacles by age 2 [206]. Zone I involvement tends to respond better to treatment with anti-VEGF agents [207].

Since laser therapy essentially ablates the retina, with central involvement, using anti-VEGF agents can help preserve vision [208]. Treatment with anti-VEGF injection is associated with fewer cases of myopia than those of laser photocoagulation. At the same time, the recurrence rate is increased in this treatment modality, particularly with zone II involvement [209, 210]. Therefore, serial assessment of patients treated with this modality is of utmost importance for the timely detection and management of recurrent disease [211]. With the currently available evidence being inconclusive in most cases, several advantages and disadvantages are cited for each treatment modality, and the treatment outcome depends on the exact pattern of retinal involvement. AI-based methods can help ophthalmologists select each patient's most effective treatment modality. This method is beneficial since clinicians do not always agree upon the best treatment modality for patients with ROP, mainly because case evaluation is usually done on a subjective basis [143, 199].

A research investigation focused on the prognostication of ROP revealed that an algorithm could reasonably predict the long-term ophthalmic results of laser photocoagulation and anti-VEGF injection in patients previously treated for ROP. The study utilized a feedforward artificial neural network that integrated an error backpropagation learning algorithm to predict visual outcomes. This prediction was based on patient birth data, their age at follow-up, and the type of treatment they received. Patients were categorized into two groups according to their prior treatments. The primary outcome measures evaluated the variance between predicted and actual visual outcomes and assessed using the normalized root mean square error [212].

Wu et al. investigation aimed to establish and validate a predictive model for the recurrence of ROP following anti-VEGF therapy, utilizing retinal images and clinical risk factors. They constructed three prediction models using ML and DL algorithms and examined 87 cases, with 21 experiencing reactivation and 66 not experiencing reactivation. The models demonstrated promising performance, with the clinical risk factor model showing AUC values of 0.80 and 0.77 in internal and external validation, respectively. The retinal image model had an average AUC of 0.82, sensitivity of 0.93, and specificity of 0.63 in internal validation. The combined model achieved specificity 0.73, AUC 0.84, and sensitivity 0.93. This predictive model could optimize treatment strategies for infants with ROP and enhance post-treatment screening plans [87].

Therefore, assessment of images and even available demographic data and the previously recognized risk factors can help design an AI-based model that predicts the most beneficial action regarding the management of each patient. Implementing such algorithms can help guide the best temporal intervals for evaluating each patient after treatment. However, no prognostic AI-based systems are currently in clinical use, and their routine implementation is accompanied by many challenges, some of which will be addressed later. An overview of the mentioned AI-based technologies used for ROP assessment is shown in Fig. 6. The table of the supplementary file summarizes the studies conducted on the application of AI in ROP. Also, Fig. 7 shows the time trend of using ML and DL algorithms in the last few years.

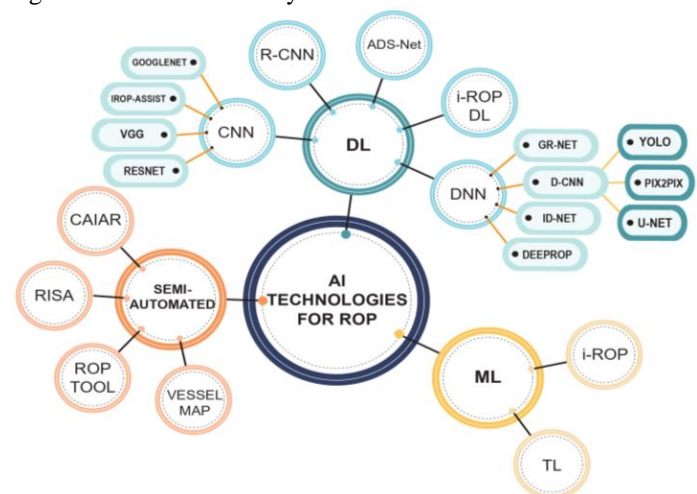

Fig. 6. An overview of AI-based techniques used for ROP detection.

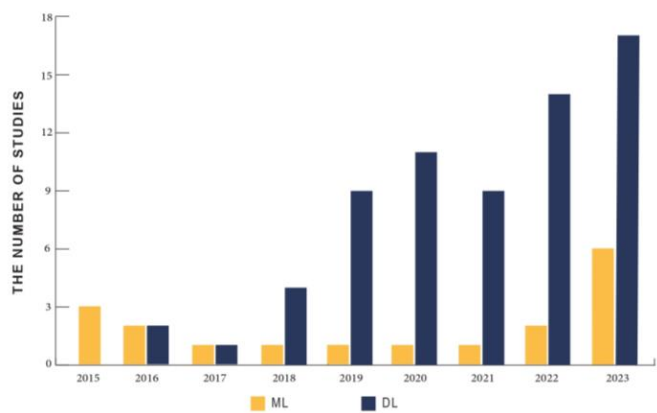

Fig. 7. Bar chart showing the number of AI studies published in ROP detection.

### E. Detabanks for Training AL Models

*FIVES (Fundus Image Vessel Segmentation):* The FIVES dataset comprises 800 high-resolution multi-disease color fundus images with pixel-level manual annotations. Medical experts standardized the annotation process through crowdsourcing, ensuring image quality. This dataset is currently the most extensive resource for retinal vessel segmentation, promising valuable contributions to its future development [213].

*STARE (Structured Analysis of the Retina):* This Project, initiated by Dr. Michael Goldbaum at the University of California, San Diego, in 1975 and funded by the U.S. National Institutes of Health, is a rich dataset contributed to by over 30 individuals from diverse fields. The dataset includes approximately 400 raw images available for download, smaller image sets for easier viewing, diagnosis codes with corresponding diagnoses, expert annotations detailing image features, blood vessel segmentation, artery and vein labeling, and optic nerve detection data. This dataset is a valuable resource for retinal image analysis and research [214].

*Kaggle:* Kaggle is a top spot for retinal image data and competitions in data science and medical imaging. These datasets help understand eye health, find diseases, and retinal image segmentation [215].

*RFMiD (Fundus Multi-disease Image Dataset):* RFMiD is divided into two parts. The RFMiD_All_Classes_Dataset includes original color fundus images, divided into training, validation, and testing sets. It consists of 3,200 images and covers 45 different types of diseases. Groundtruth labels for normal and abnormal categories are provided in CSV format. The RFMiD_Challenge_Dataset also includes original color fundus images, divided into training, validation, and testing sets. It consists of 3200 images and includes ground truth labels for 28 categories. These images were taken with three distinct fundus cameras and annotated by two experienced senior retinal specialists. RFMiD stands out as the sole publicly accessible dataset offering such a broad spectrum of retinal diseases. Its objective is to facilitate the creation of robust models for general retinal screening [216].

*DRIVE (Digital Retinal Images for Vessel Extraction):* The dataset was explicitly curated for research focused on retinal vessel segmentation. Acquired from individuals with diabetic retinopathy in the Netherlands, the dataset comprises images sourced from 400 diabetic patients aged between 25 and 90 years. Among these, 40 images were chosen randomly, with only seven exhibiting indications of mild early diabetic retinopathy. The process involved using automatic retinal map generation and branch point extraction to chronicle temporal or multimedia images and synthesize a mosaic of the retinal image.

*RITE (Retinal Images vessel Tree Extraction) database:* It collects 40 sets of retinal fundus images. It is derived from the DRIVE database and is used for comparative studies on artery and vein segmentation and classification. The database includes a vessel reference standard, a fundus photograph, and an Arteries/Veins (A/V) reference standard. The images are divided into training and test subsets with the corresponding reference standards. The A/V reference standard labels arteries in red, veins in blue, and uncertain vessels in white [217].

*MESSIDOR (Methods to evaluate segmentation and indexing techniques in retinal ophthalmology):* The MESSIDOR database consists of 1200 retinal fundus images captured using a 3CCD camera. The images have different resolutions, and 13 duplicates have been found. Inconsistencies in image grading have also been reported [218].

*REVIEW (Retinal Vessel Image set for Estimation of Widths) database:* was developed by researchers from the University of Lincoln, UK. It serves as a reference for vessel segmentation algorithms and contains 16 images with vessel width markings made by three experts. The database is segmented into four sub-databases, each comprising a specific image set. These include the high-resolution image set (HRIS) with eight images, the vascular disease image set (VDIS) containing four images, the central light reflex image set (CLRIS) featuring two images, and the kick points image set (KPIS), which consists of two images as well [219].

*ARIAS (Automated retinal image analysis system):* This system delineates blood vessels within retinal images and identifies fundamental features in digital color fundus images. The identified features encompass blood vessels, optic disc, and fovea. A proposed algorithm, the 2D matched filter response, has been introduced for blood vessel detection. Additionally, automatic recognition and localization techniques for the optic disc and fovea have been introduced and deliberated. Furthermore, a methodology for discerning left and right retinal fundus images has been put forth [220]. Table II summarizes all mentioned retinal image datasets.

TABLE II
SUMMARY OF RETINAL IMAGES DATASETS.

| Dataset | Year | Source | Image Modality | Image Type | Disease Covered | Size of Data | Resolution | Annotations |
|---|---|---|---|---|---|---|---|---|
| FIVES [213] | 2022 | China | Fundus Image | diseased retinal images | Multi-disease | 800 | High-Resolution | Retinal vessel segmentation |
| STARE [214] | 1975 | USA | Fundus Image | diseased retinal images | Multi-disease | 20 | 605 × 700 | Optic disc detection and retinal vessel segmentation |
| RFMiD [216] | 2021 | India | Fundus Image | diseased retinal images | Multi-disease | 3200 | High-Resolution | Retinal vessel segmentation |
| DRIVE [219] | 2004 | Netherlands | Fundus Image | diseased retinal images | Diabetic retinopathy | 40 | 565 × 584 | Retinal vessel segmentation and diabetic retinopathy grading |
| REVIEW [219] | 2008 | UK | Fundus Image | diseased retinal images | Multi-disease | 16 | High-Resolution | Retinal vessel segmentation |
| MESSIDOR [218] | 2013 | France | Fundus Image | diseased retinal images | Diabetic retinopathy | 1200 | 1140 X 960, 2240 X1488, and 2304 X 1536 | Retinopathy grading and risk of macular edema |
| RITE [217] | 2013 | USA | Fundus Image | diseased retinal images | N/A | 40 | High-Resolution | Retinal vessel segmentation |
| ARIAS [220] | 2011 | N/A | Fundus Image | N/A | N/A | 143 | 768 X 576 | Retinal vessels, fovea and optic disc segmentation |

## VI. Limitations and opportunities

### A. Opportunities

AI–based systems and tools have already emerged in ophthalmology as a new means of evaluation. In particular, AI has captured the attention of researchers in retinal diseases [221]. Newer AI methods, such as DL algorithms, have enhanced the healthcare system's power and accuracy in correctly detecting and managing ROP. Screening is one of the most essential key points in the ROP approach, and now, telemedicine can be used widely.

Cloud-based systems are increasingly being used and present further opportunities in the domain of AI assistants for medicine. These systems offer several advantages, such as accessibility through various devices, cost-effectiveness, scalability, and security. eClinicalWorks has introduced a cloud-based AI assistant that has enabled a family medicine clinic to see 10% more patients daily by providing swift access to concise highlights of a patient's medical history. Cloud computing is also facilitating the transition of effective and safe AI systems into mainstream healthcare, providing the computing capacity for the analysis of large amounts of data at higher speeds and lower costs compared to traditional infrastructure. These advancements demonstrate the potential of cloud-based systems in enhancing the efficiency and effectiveness of AI assistants in medicine [222, 223].

### B. Chances at better diagnosis

It is important to state that AI offers a chance at an objective evaluation for diseases that have always previously been evaluated subjectively [224]. For retinal diseases in particular, inter-personal variation of opinions among experts has been demonstrated in disease diagnosis, staging, treatment, and prognostication [122, 125, 142, 225].

### C. Technical limitation

AI systems use complex computer programs to analyze large amounts of patient information and offer suggestions for medical diagnoses. Although these systems are often very accurate, there are uncertainties about how reliable and understandable their suggestions are. Uncertainty quantification in AI means figuring out and managing these uncertainties in the suggestions made by AI models, which is important for making sure that these systems are reliable and safe to use, especially in healthcare [226].

The idea of adding uncertainty quantification to AI in healthcare has become more important recently. A review by Seoni et al. highlights how uncertainty estimation can make AI systems in healthcare more reliable and safer to use. The article underscores the potential of uncertainty quantification techniques to improve the accuracy and reliability of medical diagnoses and treatment recommendations, particularly in medical imaging applications. It also calls for further research to investigate the practical application of these techniques in real-world healthcare settings, ultimately aiming to enhance patient outcomes by effectively managing and mitigating uncertainty in clinical decision-making [227].

Even though AI models have reached human-level performance, their practical application remains restricted because they are often viewed as inscrutable. This lack of trust is a significant reason for their limited adoption in various fields, particularly healthcare [228]. Inherent in the use of AI is the need for a large number of data to be used in the training process of neural networks to produce a more comprehensive and accurate system. Although a large initial population may improve the diagnostic accuracy of the systems in use, achieving a standard objective classification system is challenging when experts have varying opinions on individual patients in many aspects of their care [122, 229].

It is also important to note that among different geographical regions, discrepancies are rather significant, especially on the thresholds for treatment or expert opinion on staging. Inherent in this method of data collection includes potential biases [230]. Therefore, systems using only local images have the limitation of being only applicable to the selected population.

Overcoming AI model application challenges necessitates enhancing transparency and interpretability, particularly through implementing explainable AI techniques. Trust-building requires rigorous validation, transparent communication, and ethical considerations. Efficient data utilization methods, such as transfer learning, help alleviate the need for extensive datasets. Collaborative guidelines and ensemble models can mitigate standardized classification complexities. Addressing regional biases involves diverse dataset curation and robust bias detection strategies. The limitation of systems dependent on local images is eased through collaborative data sharing and fostering adaptability. These solutions collectively promote responsible AI deployment, surmounting challenges and fostering inclusivity and ethical soundness in AI applications.

### D. Ethical limitations

While technical limitations exist, the most significant concern regarding the use of AI in medicine is the fact that ethical, legal, and social implications of machine-based clinical practice have not been addressed, the most relevant one of which is the potentially unfavorable effect on the patient-physician relationship [231]. With the use of digital systems, privacy and data ownership issues appear, along with informed patient consent for using their medical information in the development of the said systems [232]. Therefore, strong protections, such as encrypting data and ensuring informed consent, help keep sensitive information safe.

Ethical issues are especially prominent when the AI-based system in question is concerned with prognosticating a particular disease process since the implication of the system here tends to be futuristic, and patients often experience reluctance to accept the automated verdict for their care [233]. Clinicians also tend to be skeptical about using AI in their clinical decision-making since the issue of accountability for the decision made via AI systems is poorly defined; therefore, before AI-based systems get into routine clinical use for treatment decision-making and prognosis, specific guidelines need to be implemented to define the accountability profiles in cases where misdiagnoses occur [221, 234, 235].

### E. Financial issue

The extensive implementation of AI solutions in healthcare settings may be impeded by financial issues, especially in areas or institutions with low funding. Healthcare organizations may face significant financial difficulties with the initial investment in AI technology, which includes developing AI algorithms and obtaining advanced imaging equipment. Moreover, ongoing expenses for system maintenance, updates, and staff training further contribute to the financial challenges associated with AI integration.

Collaboration between governmental agencies, private companies, and academic institutions is necessary to address the financial support issue. Allocating funding for AI research and implementation in ROP management can facilitate technological advancements and ensure accessibility to a broader range of healthcare facilities. Additionally, fostering partnerships between healthcare providers and technology developers can help distribute the financial burden and promote sustainable, cost-effective solutions. A summary of the challenges in employing AI to ROP assessment and suggested solutions to overcome them are shown in Fig. 8.

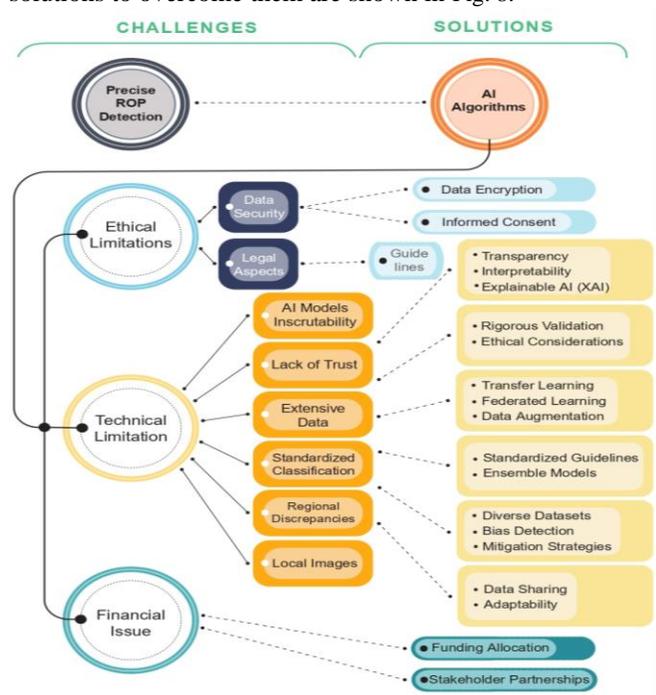

Fig. 8. Challenges and suggested solutions in employing AI in ROP.

## VII. CONCLUSION

As the evolution of AI applications in healthcare is remarkable, we have also concluded this progress in ROP management in recent years. Early semi-automated systems aim to quantify retinal features to aid in diagnosing plus disease but face limitations in accuracy and generalizability. Four critical computer-based ROP diagnostic systems - ROPTool, RISA, VesselMap, and CAIAR - extract retinal image features and use manual or semi-automated methods to assess dilation and tortuosity, showing varying diagnostic alignment with clinical ROP diagnoses. Newer models like i-ROP have two critical advantages compared to previous models: it prevents false classification of normal eyes as plus-disease and avoids misclassifying plus-disease eyes as usual. Recent advancements have seen the rise of DL models, particularly CNN and its variants, which have demonstrated remarkable efficacy in identifying specific ROP-related features with accuracy comparable to human experts. The i-ROP-DL system, integrating two CNNs for vessel segmentation and classification, has emerged as a promising method for precisely detecting plus disease. The potential of AI in ROP management lies in its ability to provide more objective evaluation, enhance diagnostic accuracy, and facilitate personalized treatment selection for at-risk infants. By automating ROP screening and diagnosis, AI-based systems can reduce subjectivity, enhance efficiency, and improve patient outcomes. However, while AI offers exciting possibilities, challenges remain in building diverse data sets to ensure the robustness and generalizability of AI models. In light of the transformative potential of AI in ROP management, further research, validation, and collaborative efforts are necessary to integrate AI technologies into clinical practice effectively. By harnessing the power of AI, we can revolutionize ROP diagnosis and management, ensuring timely and accurate intervention to safeguard the precious vision of premature infants. With continued progress, AI-based systems hold the key to a brighter future for ROP and other retinal diseases.

TABLE OF SUPPLEMENTARY
SUMMARY OF STUDIES CONDUCTED REGARDING THE APPLICATION OF AI IN ROP

| Author | Year | Participants | Methods | Results |
|---|---|---|---|---|
| Capowski et al. [29] | 1995 | 20 images | Assessed retinal vessel tortuosity using sequential images. Sensitive to ROP changes, computer processed, | Created indicator using spatial frequency, noticing the distinct frequency of tortuous arteries in plus disease |
| Heneghan et al. [30] | 2002 | 11 ROP, 9 non-ROP participants | created a software called Vessel Finder, which computed vascular dilation and tortuosity | Using vessel width and tortuosity, the sensitivity and specificity of the test improved to 82% and 75% |
| Rabinowitz et al. [31] | 2007 | Digital fundus images from 78 eye | assessed a semiautomated program known as VesselMap | predicted severe ROP, with AUORC scores ranging from 0.75 to 0.94 |
| Chiang et al. [32] | 2007 | 67 infants | Uploaded data to web system, 3 specialists diagnosed ROP levels from none to treatment-requiring. | Telemedicine systems using nurse-captured retinal images improve ROP management, especially at later PMAs. |
| Chiang et al. [33] | 2007 | 34 wide-angle retinal images | Assessing Agreement and Accuracy in Plus Disease Diagnosis for ROP Experts and Comparing with Computer-based Analysis | Among individual computer system parameters compared to the reference standard, venular IC had the highest AUC (0.853). Among linear combinations of parameters, the combination of arteriolar IC, arteriolar TI, venular IC, venular diameter, and venular TI had the highest AUC (0.967) |
| Wilson et al. [34] | 2008 | 75 vessels on 10 retinal images | CAIAR program measures retinal vessel features from images, validated with computer models, and compared to expert ophthalmologist grading using real images | A moderate correlation was found in 10 of the 14 tortuosity output calculations (Spearman rho = 0.618-0.673). Width was less well correlated (rho = 0.415) |
| A Ray et al. [35] | 2008 | 358 infants' medical data | Applying ML methods for the prediction of ROP progression in newborns | highest accuracy reported 84.36% |
| Wallace et al. [36] | 2009 | 20 RetCam images from premature infants | use ROPtool for measuring and analyzing both tortuosity and dilation | AUCs for plus disease and pre-plus disease were 0.93 and 0.90, respectively. ROPtool had 89% sensitivity and 83% specificity for identifying appropriate dilatation for plus conditions |
| Wittenberg et al. [37] | 2012 | A dataset of retinal images | examined the four primary computer-based diagnostic systems for ROP: ROPTool, RISA, VesselMap, and CAIAR | ROPtool (AU ROC: plus tortuosity 0.95, plus dilation 0.87), RISA (AU ROC: arteriolar TI 0.71, venular diameter 0.82), VesselMap (AU ROC: arteriolar dilation 0.75, venular dilation 0.96), and CAIAR (AU ROC: arteriole tortuosity 0.92, venular dilation 0.91) |
| Wilson et al.[38] | 2012 | digital images from 75 infants | Images were standardized to 240-pixel diameter and then assessed using semiautomated software CAIAR for retinal vessel width and tortuosity measurements. | In CAIAR, differentiating arterioles from venules isn't required to distinguish ROP stages. Tortuosity seems more reliable than width to differentiate eyes with and without ROP. |
| A.Cansizoglu et al. [39] | 2015 | 77 wide-angle retinal images | A computer technique was created to extract tortuosity and dilatation properties from arteries and veins by cropping pictures that had been manually divided into various shapes and sizes | point-based tortuosity characteristics of acceleration and curvature were shown to outperform the segment-based or dilation measures |
| Cansizoglu et al [40] | 2015 | 77 wide-angle retinal images | employed an SVM to determine the optimal combination of features and field of view that correlated best with expert diagnosis | achieved its highest accuracy (95%) when it incorporated vascular tortuosity measurements from both arteries and veins |
| Bolón-Canedo et al. [41] | 2015 | 34 retinal images | Feature selection and an ML approach for evaluating the inter-expert viability in ROP | Improved ML system performance by detecting the most relevant features |
| Bolón-Canedo et al. [42] | 2015 | 77 fundus images | Used a Gaussian Mixture Model (GMM) for feature extraction and vessel segmentation for ROP diagnosing | The model Achieved an accuracy of 90% |
| Worrall et al. [43] | 2016 | 1459 images from 35 patients | explored two approaches to assist clinicians in detecting ROP using CNN | The system's image detection classifier was almost as accurate (92%) as the human medical professionals |
| Abbey et al. [44] | 2016 | 335 fundus images | ROPtool was used for quantitative analysis, comparing its grades with expert panel grades via ROC curve and compared with the lay reader. Inter-reader grades cross-tabulated | a sensitivity of 71-86% for plus detection of plus diseases, and a sensitivity of 91% for any vascular abnormality |
| Campbell et al. [45] | 2016 | 77 digital fundus images | Used i-ROP to identify the vascular characteristics on which physicians rely for the diagnosis of plus disease | A CBIA system might potentially achieve ROP expert-level performance using manually segmented images. |
| E. R. Rajkumar et al.[46] | 2016 | 56 fundus images | Two classifiers, miGraph and citation-kNN were used for the classification of ROP. | Accuracy of %89.64 for Citation-kNN and %95 for miGraph |
| Pour et al. [47] | 2017 | 87 fundus pictures | aimed to provide a unique software program with new algorithms that assess vascular tortuosity and dilatation in fundal images | The accuracy rate for this study was 80.15 % |
| Kalpathy-Cramer et al. [48] | 2017 | 195 fundus images | Three different DL approaches were compared together in the classification of ROP disease. | U-network-based architecture yielded an AUC of 0.97, |
| S Ostmo et al. [49] | 2018 | 100 fundus images | i-ROP ASSIST, a supervised ML algorithm, was used to develop a quantitative severity scale for ROP | Achieved a sensitivity of 93.9 and specificity of 98.2 % in detection of ROP |
| Brown et al. [50] | 2018 | a data set of 5511 retinal images | Used CNN for segmenting retinal arteries and veins, detecting additional pathology like plus disease | their CNN method obtains AUC values of 0.94 and 0.98, respectively |
| Y. Zhang et al. [51] | 2018 | 17801 fundus images | Three different DNN classifiers were applied for an automated screening system for ROP. | The final DNN model reached an accuracy of 98.8% and 93% precision. |
| Wang et al. [52] | 2018 | A large dataset of fundus images | Two DNN models, i.e., Gr-Net and Id-Net, were considered for grading and identification tasks, respectively | a sensitivity of 84.91 and specificity of 96.90% in the DeepROP for ROP identification, whereas the equivalent measures for ROP grading were 93.33% and 73.63%, respectively. |
| M. F. Chiang et al. [53] | 2018 | 31 wide-angle retinal images | Used i-ROP DL for analyzing the most relevant features of retinal images in ROP detection | Some features recognized as most relevant in ROP detection include dilation and tortuosity of both arteries and veins and the location of central retinal vessels. |
| Mulay et al. [54] | 2019 | 220 images of 45 infants | developed a regional CNN-based algorithm to identify ridges to better classify Stage 2 ROP | the system achieved a detection accuracy rate of 0.88, proving that the early-stage ROP can be reliably detected using DL with pre-processing through image normalization |
| S. Taylor et al. [55] | 2019 | 5255 clinical examinations of 871 premature infants | A DL-based program (i-ROP DL) was used to assess the ROP progression and vascular severity score | The ROP vascular severity score is associated with disease category and clinical progression in premature infants. Automated image analysis can identify high-risk infants for TR-ROP. |
| T. K. Redd et al. [56] | 2019 | 870 infants' fundus images and examination | Assessed the i-ROP DL system potential for screening and detection of ROP | The system reached an AUC od 96% for detecting type 1 ROP |
| T. Redd et al. [57] | 2019 | 568 fundus images | A DL-based system trained for assessing the ROP epidemiologic surveillance | Significant disease burden differences were identified among Aravind system hospitals, independent of known ROP risk factors. This has the potential for evaluating primary prevention variations in LMIC hospitals. |
| Z. Tan et al. [58] | 2019 | 6974 fundus images | A DL-based tool named ROP.AI trained for automatic detection of ROP | Achieved an AUC of 99.3%, accuracy of 97.3% within the 20% hold-out test set |
| Wang Ji et al. [59] | 2019 | 1464 fundus photographs | Diagnosing ROP by DL approaches in the manually labeled retinal images from ROP and non-ROP infants and grading | The combination of DL and human-machine collaboration achieved a good level of success in diagnosing stage disease, with an accuracy rate of 94.08% and a Kappa value of 0.880 |
| K. N. Smith et al. [60] | 2019 | 7264 clinical data | Images were analyzed using i-ROP DL and assigned a vascular severity score from 1-9. Demographic data, systemic comorbidities, and post-menstrual age at peak disease severity were evaluated for each category | Infants with TR ROP who developed APROP had lower birth weight and gestational age, reached peak severity at an earlier post-menstrual age, and showed the correlation between the mean i-ROP vascular severity score and RSD-based ROP severity categories. |
| J. Mao et al. [61] | 2019 | Three distinct datasets of retinal images | A deep CNN algorithm was applied for vessel segmentation and classification of ROP and non-ROP conditions. | The study reported a sensitivity of 95.1% with a specificity of 97.8% for diagnosing ROP. |
| G. Chen et al. [62] | 2019 | 7330 retinal images | They combined segmentation and staging using FCN and MIL to achieve integrated ROP staging and lesion localization. | The proposed network achieved an AUC of 0.93 and an accuracy of 92.25% |
| J. P. Campbell et al. [63] | 2020 | 6354 Retcam images | Used a DL-based vascular scale for ROP staging (i-ROP cohort) | The vascular severity score increased with increasing stage of disease in zone I (top left, P<0.001) and zone II (top right, P<0.001), and with increasing extent of stage 3 in zone I (bottom left, P=0.03) and zone II (bottom right, P<0.001) |
| K. N. Bellsmith et al. [64] | 2020 | 5945 clinical eye examinations | Used DL-based methods for grading TR-ROP | AP-ROP infants had lower birth weight and GA compared to those without AP-ROP. They reached peak severity earlier |
| | | | | (34.7 weeks vs. 36.9 weeks; P < 0.001) and had a higher mean vascular severity score (8.79 vs. 7.19; P < 0.001) |
| Y. Luo et al. [65] | 2020 | A dataset of retinal images | They used a U-net for segmenting blood vessels and the optic disc and integrated four pathological features with the network's feature vectors to create a fusion DNN. | Incorporating these features into the neural network greatly improved its performance compared to the original network. |
| J. Chen et al. [66] | 2020 | 4441 retinal images | An automated CNN network assessed for the predictive capability of ROP | The model achieved an AUC-ROC of 0.98 |
| V. M. Yildiz et al. [67] | 2020 | two datasets consist of 5512 and 100 retinal images | ROP-relevant features such as tortuosity and dilation measures were extracted. These features were used in classifiers (logistic regression, SVM, and neural networks) to assess the severity score of the input | The algorithms achieved an AUC of 0.88 and 0.99, respectively, in two distinct datasets |
| Y. Tong et al. [68] | 2020 | 36,231 fundus images | Two DL methods, CNN and Faster-RCNN were trained to classification and detection of ROP | The system achieved an accuracy of 90.3% for classification and 95.7% for detecting the ROP stage. |
| D. Lepore et al. [69] | 2020 | 835 fluorescein angiography images | Investigating the applicability of a CNN algorithm for ROP management via fluorescein angiography images | Accuracy of 88% and AUC of 0.91 were reported. |
| Xin Guo et al. [70] | 2020 | 2 different datasets of retinal images | Compared 3 models based on CNN architecture for the classification of ROP | The CNN-based models achieved an accuracy of 0.93, 0.8948, and 0.98 in the different 3 model |
| A. Ding et al. [71] | 2020 | A dataset of retinal images | They trained a CNN program for the staging of ROP in 1-3 stages | The architectures yielded an accuracy of 0.67, 0.54, and 0.47 in the hybrid, only classifier and object segmentation, respectively |
| N. Valikodath et al. [72] | 2020 | 15467 eye examinations | i-ROP DL system was applied for analyzing the images and determining the severity score of ROP | When the cutoff score was set at 3, the sensitivity for TR-ROP was 86%, while the specificity was 66%. |
| S. Ramachandran et al. [73] | 2020 | 10 different datasets | They proposed a DNN-based framework for OD localization, utilizing YOLOv3, a fully convolutional neural network pipeline. | Overall accuracy was reported as 99.25% for ROP detection. |
| Peng et al. [74] | 2021 | 635 retinal fundus images | A three-stream parallel framework including ResNet18, DenseNet121, and EfficientNetB2 used for ROP staging | achieved 0.9055 for weighted recall, 0.9092 for weighted precision, 0.9043 for weighted F1 score, 0.9827 for accuracy with 1 (ACC1) and 0.9786 for Kappa, respectively |
| A. Vinekar et al. [75] | 2021 | 42,641retinal images from the tele-ROP screening program of India | Developed and validated a DL screening tool on retinal images obtained through a tele-ophthalmology platform for ROP | Reported a sensitivity of 95.7% for test A and 97.8% for B, also a specificity of 99.6% for A and 68.3% for B |
| R. Agrawal et al. [76] | 2021 | 10,000 Retcam images | A DL method (U-Net) was applied to automatically ROP detection | Reached An AUC of 0.95 for optic disc discrimination and 1 for blood vessel segmentation, and 0.95 for training |
| Chen B.A. et al. [77] | 2021 | 5943 fundus images (Retcam) | Two CNN model trained for ROP staging in newborns | The North American model showed an AUROC of 0.99, and Nepali-trained showed 0.97 AUROC |
| O.Attallah et al. [78] | 2021 | 8090 retinal images from ROP, 9711 from normal eyes | suggests a dependable automated diagnostic tool named DIAROP, which utilizes DL techniques to aid in ophthalmologic ROP diagnosis | DIAROP attained an accuracy rate of 93.2% along with an impressive AUC of 0.98 |
| S. Ramachandran et al. [79] | 2021 | 289 retinal images | DL architecture was used for ROP detection and bound boxes for vascularization of the retina. | A sensitivity of 99% and specificity of 98% were reported for DL algorithm performance. |
| J. Zhang et al. [80] | 2021 | 521,586 objects | A meta-analysis of applied DL models for ROP detection | An AUC for combined validation and test datasets were 0.984, for the validation and test dataset were 0.977, and in the subgroup analysis of ROP were 0.99 |
| J. P. Campbell et al. [81] | 2021 | 499 fundus images | DL-derived vascular severity score evaluated for ROP staging | zone I had a higher vascular severity score than zones II and III (P<.001). Zone I also had a higher severity score than zone II for a given number of clock hours of stage 3 (P=.03 in zone I and P<.001 in zone II) |
| A. S. Coyner et al. [82] | 2021 | 1579 fundus images | ML-derived (ElasticNet) method used for TR-ROP detection | The area under the precision-recall curve was 0.35 ± 0.11, and a sensitivity of 100% was detected |
| Scruggs et al. [83] | 2022 | 13 neonates | wider-angle OCT records to assist in the identification, staging, and analysis of peripheral aspects of ROP | The earliest visible indicators of proliferation are small groups of neo-vessels seen behind the ridge |
| Peng et al. [84] | 2022 | two datasets, the first one includes 7396 fundus images, and the second one includes 1337 fundus images | a new deep supervision-based network (ADS-Net) is projected to detect ROP and determine 3-level ROP grading | ADS-Net accomplished 0.9552 and 0.9037 in ROP screening and grading for the Kappa index |
| R. R. Struyven et al. [85] | 2022 | 6620 retinal images | Two models were trained for ROP diagnosis: a custom DenseNet model and a Google Cloud AutoML Vision model trained through the online interface without coding. | The custom coding model achieved an AUC of 0.941 and 95.2% accuracy for normal versus diseased classification, while the AutoML model achieved an AUC of 0.95 and 96.7% accuracy for the same classification. |
| Peng Li et al. [86] | 2022 | Three datasets (18827) of retinal images | A CNN trained for diagnosing the ROP stages between I and III | Reported an average sensitivity of 91.6% and specificity of 98.56 for three stages of discrimination |
| Wu et al. [87] | 2022 | 87 infants | create and validate a prediction model for ROP reactivation after anti-IGF-1 therapy | The clinical risk factor model showing AUC values of 0.80 and 0.77 in internal and external validation |
| Campbell et al. [88] | 2022 | Two distinct datasets of 30 fundus pictures | The retinal images were labeled with a DL-derived score by the iROP-DL algorithm for the severity of vascular disease from 1 to 9 | The vascular severity score was significantly associated with mode plus label (P < 0.001) and with the ophthalmoscopic diagnosis of the stage in the same eyes (P < 0.001) and correlated well with the average disease severity (CC = 0.90) |
| Q. Wu et al. [89] | 2022 | 815 retinal images | Used a DL approach to predict the occurrence and severity of ROP in newborns | Reported an AUC of 90% and 87% for predicting the occurrence and severity of ROP |
| Y. P. Huang et al. [90] | 2022 | 176 fundus images | Computed the temporal artery angle (TAA) and temporal vein angle (TVA) within the temporal quadrant to detect ROP | TAA and TVA decreased, while TAW and TVW increased with higher ROP severity (all P < 0.0001). Positive correlations were seen between TAA–TVA and TAW–TVW (both P < 0.0001). TAA showed a negative correlation with TAW (r = −0.162, P = 0.0314), potentially aiding ROP detection by ophthalmologists. |
| L. Jie et al. [91] | 2022 | 54,626 fundus images | A DL approach analyzed images, and a quantitative method, DeepROP, was applied to assess retinal vascular abnormality. | DeepROP score showed an AUC of 0.981 for type 1 and 0.986 for type 2 detection. |
| A. S. Coyner et al. [92] | 2022 | 5842 retinal images | RVMs were extracted from retinal images and used to train PGANs. CNNs were then trained on real or synthetic RVMs to detect plus disease. | The CNN showed an AUC of 0.97 on synthetic RVM and 0.93 on real RVM |
| L. Ju et al. [93] | 2022 | A real clinical dataset | proposed a new semiautomated DL framework for ROP staging | The model presented an AUC of 0.863 and an accuracy of 0.789 for ROP staging |
| A. Subramaniam et al. [94] | 2022 | 441 fundus images | A preliminary binary classifier using GoogLeNet was developed to distinguish between plus and no plus diseases. Smartphone images were pre-processed, including vessel enhancement, size normalization, and reasonable augmentation. | Reached an accuracy of 0.96 for limited data |
| R. Agrawal et al. [95] | 2022 | Two distinct datasets | Three DL models, U-Net, AG U-Net, and SE U-Net, were used for fundus image segmentation for ROP detection. | An AUC of 0.94 was reported for all three networks. |
| Z. Luo et al. [96] | 2022 | 900 color fundus images | A combination of DL algorithms and telemedicine systems was applied to develop a screening program for ROP. | The program reached an AUC of 60% and an ACC of 75% |
| C. Lu et al. [97] | 2022 | 5255 wide-angle retinal images | DL models were trained, validated, and tested on labeled retinal images to detect plus, pre-plus, or no plus disease. The labels were determined by 3 image-based ROP graders and the clinical diagnosis, serving as a reference standard diagnosis. | Out of the 7 local models trained, 4 (57%) performed worse than the FL models. The performance of the local models was positively correlated with label agreement, total number of plus cases, and overall training set size. |
| V. Kumar et al. [98] | 2023 | 439 preterm neonatal retinal images | Used a deep CNN with YOLO-v5 for OD detection and Pix2Pix, U-Net, or a deep CNN for retinal blood vessel segmentation. | OD detection reached 98.94% accuracy (IoU 0.5), and blood vessel segmentation scored 96.69% accuracy, with a Dice coefficient of 0.60 to 0.64. Zone-1 ROP diagnosis achieved 88.23% precision. Our method presents a promising avenue for precise ROP screening and diagnosis. |
| W. C. Lin [99] | 2023 | 230 infants | Aimed to assess the potential association between time-series oxygen data within the EHR and the occurrence of type 2 ROP or ROP requiring treatment (TR-ROP). | LSTM models outperformed best ML models (SVM with GA and 3 average FiO2 features) and SVM models on GA, with a mean AUROC of 0.89±0.04 versus 0.86±0.05 and 0.83±0.04. They also achieved the highest F1 score (0.85±0.06), followed by SVM with 4 variables (0.82±0.07) and SVM with GA alone (0.80±0.06). |
| N. Salih et al. [100] | 2023 | 1365 fundus images | Four distinct models for transfer learning and CNN architecture were trained, including VGG-19, ResNet-50, and EfficientNetB5, for the recognition of ROP zones in premature infants. | The voting classifier achieved an aggregate accuracy of 88.82%. Additionally, EfficientNetB5 demonstrated superior performance in accuracy compared to alternative models, boasting a rate of 87.27%. |
| K. M. Jemshi et al. [101] | 2023 | 178 retinal images | They presented an artificial neural network architecture with optimized features to fulfill the critical requirement of a Plus disease classifier, eliminating false negatives in an ROP screening system. | By incorporating Curvelet transform energy coefficients and vascular features, the system achieved 96% accuracy, 93% specificity, and 100% sensitivity. |
| A. T. Legocki et al. [102] | 2023 | 167 OCT images | developed multivariate models utilizing demographic factors and OCT imaging results to detect early referral-worthy ROP (including referral-warranted ROP and pre-plus disease). | The AUC was 0.94 for the generalized linear mixed model and 0.83 for the ML model; key variables: birth weight, Opacity Ratio, vessel factors, Simplified models: Birth weight + gestation: AUC 0.68. Imaging alone: AUC 0.88 |
| K. Wagner et al. [103] | 2023 | 1370 retinal images | Used two DL methods: Bespoke and code-free deep learning models (CFDL) for discrimination between ROP and healthy eyes | Reached an AUC Of 0·989 for CFDL and 0·986 for Bespoke model |
| R. Sankari et al. [104] | 2023 | 4000 fundus images | A CNN network (RF) designed for automated diagnosis of ROP | The hybrid model reached an accuracy of 94.5% |
| R. Sankari et al. [105] | 2023 | 200 fundus images | Four ML classifiers (Quantum SVM classifier REP tree, LogitBoost, and K-Star) for ROP classification | Achieved an accuracy of 86.7%, 75%, 76.5% and 74% respectively |
| A. Subramaniam et al. [106] | 2023 | Cell phone images harmonized | A DL-based binary classifier derived from GoogleNet was applied for the classification of plus and no plus conditions. | Achieved an AUC of 0.975 on a limited dataset |
| S. Rahim et al. [107] | 2023 | A dataset of fundus images | New fundus pre-processing methods were applied to pre-trained transfer learning frameworks to create hybrid models. | Achieved high accuracy rates of 97.65% for Plus disease, 89.44% for Stage, and 90.24% for Zones |
| X. Deng et al. [108] | 2023 | 289 retinal images from ROP | A DL network segmented retinal vessels and the OD, allowing automatic evaluation of vascular morphological characteristics. | It exhibited a sensitivity of 72% and a specificity of 99% |
| Sonja K. Eilts et al. [109] | 2023 | 1046 images from 19 ROP infants | Evaluated fundus images within the CARE-ROP trial via an AI algorithm, then assigned a severity score (VSS) | 19 infants with ROP showed decreased VSS from 6.7 to 2.7 at week 1 and 2.9 at week 4. Infants needing retreatment had higher baseline VSS (6.5). Higher baseline VSS correlated with earlier retreatment (r = -0.9997; P < .001). |
| M. A. de Campos-Stairiker et al. [110] | 2023 | 3093 images and clinical data from ROP neonates | Compared eye conditions (type 2 or treatment-requiring ROP) and AI-derived vascular severity scores in two time periods for all district babies during initial tele-retinal screening | Type 2 or worse and TR-ROP decreased: 60.9% to 17.1% (P < 0.001) and 16.8% to 5.1% (P < 0.001) in matched babies. Median VSS in the population decreased from 2.9 to 2.4 (P < 0.001). |
| Y. Liu et al. [111] | 2023 | 24,495 RetCam images from 651 preterm infants | A CNN identified ROP, severe ROP, and treatment modalities, including retinal laser or injections, and is compared to ophthalmologists. | CNN performance in ROP tasks (AUC 0.9531), severe ROP (AUC 0.9132), and treatment modalities (AUC 0.9360) with 92.0% accuracy in external validation, surpassing four ophthalmologists |
| B. Young et al. [112] | 2023 | Retcam images from 159 preterm infants | Smartphone images trained ResNet18 for binary classification (normal vs. pre-plus/plus disease), predicting referral warranted (RW) and TR-ROP at the patient level. | The AI system achieved 100% sensitivity for TR-ROP with 58.6% specificity and 80.0% sensitivity for RW-ROP with 59.3% specificity. |
| O. Attallah et al. [117] | 2023 | 2 different datasets of fundus images from preterm infants | Proposing GabROP analyzes fundus images, generates GW sets, and trains three CNNs independently. Textural-spectral-temporal representation is created using DWT and features from original fundus images | GabROP proves accurate and efficient for ophthalmologists with an AUC of 0.98, outperforming recent ROP diagnostic techniques |
| E. K. Yenice et al. [113] | 2023 | Clinical data from 640 preterm twin pairs | Variables used to develop the ROP prediction model. ML for training, validated with 10-fold cross-validation | Decision Tree detected ROP with 71% sensitivity and 80% specificity in CV, while Multi-Layer Perceptron showed 70% sensitivity and specificity. X-Tree and RF achieved 84% and 80% specificity for ROP prediction, respectively. |
| A. S. Coyner et al. [114] | 2023 | 4095 fundus images | A U-Net CNN segmented arteries and veins in RFIs into grayscale RVMs, which were thresholded, binarized, or skeletonized. CNNs trained with SRR labels on color RFIs and various forms of RVMs | Raw RVMs were nearly as informative as color RFIs (image-level AUC-PR 0.938; infant-level AUC-PR 0.995). CNNs learned to distinguish RFIs or RVMs from Black or White infants despite color, brightness, or width variations in vessel segmentation. |
| K. L. Nisha et al. [115] | 2023 | A dataset of retinal images | Individual "ground truth" prepared, unique ML classifier built for each person, facilitating identification of their most crucial features | Trainees showed 88.4% and 86.2% repeatability, while the expert demonstrated 92.1%. Notably, commonly used features varied between the expert and trainees, contributing to the observed variability |
| D. P. Rao et al. [116] | 2023 | 227,326 wide-field retinal images | The tele-ROP program dataset was split into train, validation, and test sets. A binary classifier for ROP (Stages 1–3) utilized | Sensitivity and specificity to detect ROP were 91.46% and 91.22%, respectively, AUROC was 0.970. |

\*ROP: retinopathy of prematurity, CNN: convolutional neural network, DL: deep learning, ML: machine learning, AUC: area under the curve, OCT: optical coherence tomography, TAA: temporal artery angle, TVA: temporal vein angle, CAIAR: computer-aided image analysis of the retina, SVM: support vector machine, FCN: fully convolutional network, LSTM: long short-term memory, OD: optic disc, TAW: temporal artery width, TVW: temporal vein width, IC: integrated curvature, RISA: Retinal Image Multi-Scale Analysis, APROP: Aggressive posterior retinopathy of prematurity, TR-ROP: treatment requiring ROP, REP: Reduced Error Pruning, LMIC: low- and middle-income countries, AP-ROP: Aggressive posterior retinopathy of prematurity, RVM: retinal vessels map, RFI: Retinal Fundus Image, VSS: vascular severity score.